\documentclass[prb,twocolumn,showpacs,amsmath,amssymb,superscriptaddress,bibnotes]{revtex4-2}
\usepackage{graphicx}
\usepackage{color}
\usepackage{braket}
\usepackage{amsmath}
\usepackage{amsfonts}
\usepackage[caption=false]{subfig}
\usepackage{natbib}
\usepackage{bbold}
\usepackage{verbatim}

\usepackage{amsmath,amssymb,mathrsfs}
\usepackage{bbm}
\usepackage[dvipsnames]{xcolor}

\date{\today}

\begin{document}
	
	\title{RKKY interaction in one-dimensional flat-band lattices}
	
	\author{Katharina Laubscher}
	\thanks{These authors contributed equally to this work.}
	\affiliation{Department of Physics, University of Basel, Klingelbergstrasse 82, CH-4056 Basel, Switzerland}
	\author{Clara S. Weber}
	\thanks{These authors contributed equally to this work.}
	\affiliation{Institut f\"ur Theorie der Statistischen Physik, RWTH Aachen University and JARA - Fundamentals of Future Information Technology, D-52056 Aachen, Germany}
	\affiliation{Department of Physics and Astronomy, University of Pennsylvania, Philadelphia, Pennsylvania 19104, USA}
	\author{Maximilian H\"{u}nenberger}
	\affiliation{Department of Physics, University of Basel, Klingelbergstrasse 82, CH-4056 Basel, Switzerland}
	\author{Herbert Schoeller}
	\affiliation{Institut f\"ur Theorie der Statistischen Physik, RWTH Aachen, 
		52056 Aachen, Germany and JARA - Fundamentals of Future Information Technology}
	\author{Dante M. Kennes}
	\affiliation{Institut f\"ur Theorie der Statistischen Physik, RWTH Aachen, 
		52056 Aachen, Germany and JARA - Fundamentals of Future Information Technology}
	\affiliation{Max Planck Institute for the Structure and Dynamics of Matter, Center for Free Electron Laser Science, 22761 Hamburg, Germany}
	\author{Daniel Loss}
	\affiliation{Department of Physics, University of Basel, Klingelbergstrasse 82, CH-4056 Basel, Switzerland}
	\author{Jelena Klinovaja}
	\affiliation{Department of Physics, University of Basel, Klingelbergstrasse 82, CH-4056 Basel, Switzerland}
	
	\begin{abstract}
		We study the Ruderman-Kittel-Kasuya-Yosida (RKKY) interaction between two classical magnetic impurities in one-dimensional lattice models with flat bands. As two representative examples, we pick the stub lattice and the diamond lattice at half filling. We first calculate the exact RKKY interaction numerically and then compare our data to results obtained via different analytical techniques. In both our examples, we find that the RKKY interaction exhibits peculiar features that can directly be traced back to the presence of a flat band in the energy spectrum. Importantly, these features are not captured by the conventional RKKY approximation based on non-degenerate perturbation theory. Instead, we find that degenerate perturbation theory correctly reproduces our exact results if there is an energy gap between the flat and the dispersive bands, while a nonperturbative approach becomes necessary in the absence of a gap.
	\end{abstract}
	
	\maketitle

\section{Introduction}
Magnetic impurities embedded in a host material can interact indirectly by coupling to the electron spin density of the host. This so-called Ruderman-Kittel-Kasuya-Yosida (RKKY) interaction~\cite{Ruderman1954,Kasuya1956,Yosida1957} can result in a magnetic ordering of the impurity spins, leading to a wide range of interesting phenomena with potential applications in the fields of spintronics~\cite{Bruno1991,Bruno1992}, spin-based quantum computation~\cite{Craig2004,Glazman2004,Usaj2005,Simon2005,Yang2016}, or engineered topological superconductivity~\cite{Pientka2013,Braunecker2013,Klinovaja2013b,Vazifeh2013,Pientka2014,Kim2014,Braunecker2015,Hsu2015,Schecter2016,Pawlak2016,Pawlak2019}. The exact form of the RKKY interaction depends on the properties---in particular, the band structure---of the underlying host material and has been extensively studied for various types of systems~\cite{Zyuzin1986,Poilblanc1994,Balatsky1995,Galitski2002,Imamura2004,Saremi2007,Hwang2008,Gao2009,Braunecker2009,Liu2009,Garate2010,Black-Schaffer2010,Black-Schaffer2010b,Braunecker2010,Chesi2010,Abanin2011,Zhu2011,Sherafati2011,Kogan2011,Klinovaja2013,Power2013,Zyuzin2014,Yao2014,Efimkin2014,Schecter2015,Tsvelik2017,Kurilovich2017,Hsu2017,Hsu2018,Legg2019,Ovando2019,Deb2021,Laubscher2022}.
	
	\begin{figure}[bt]
		\centering
		\includegraphics[width=0.95\columnwidth]{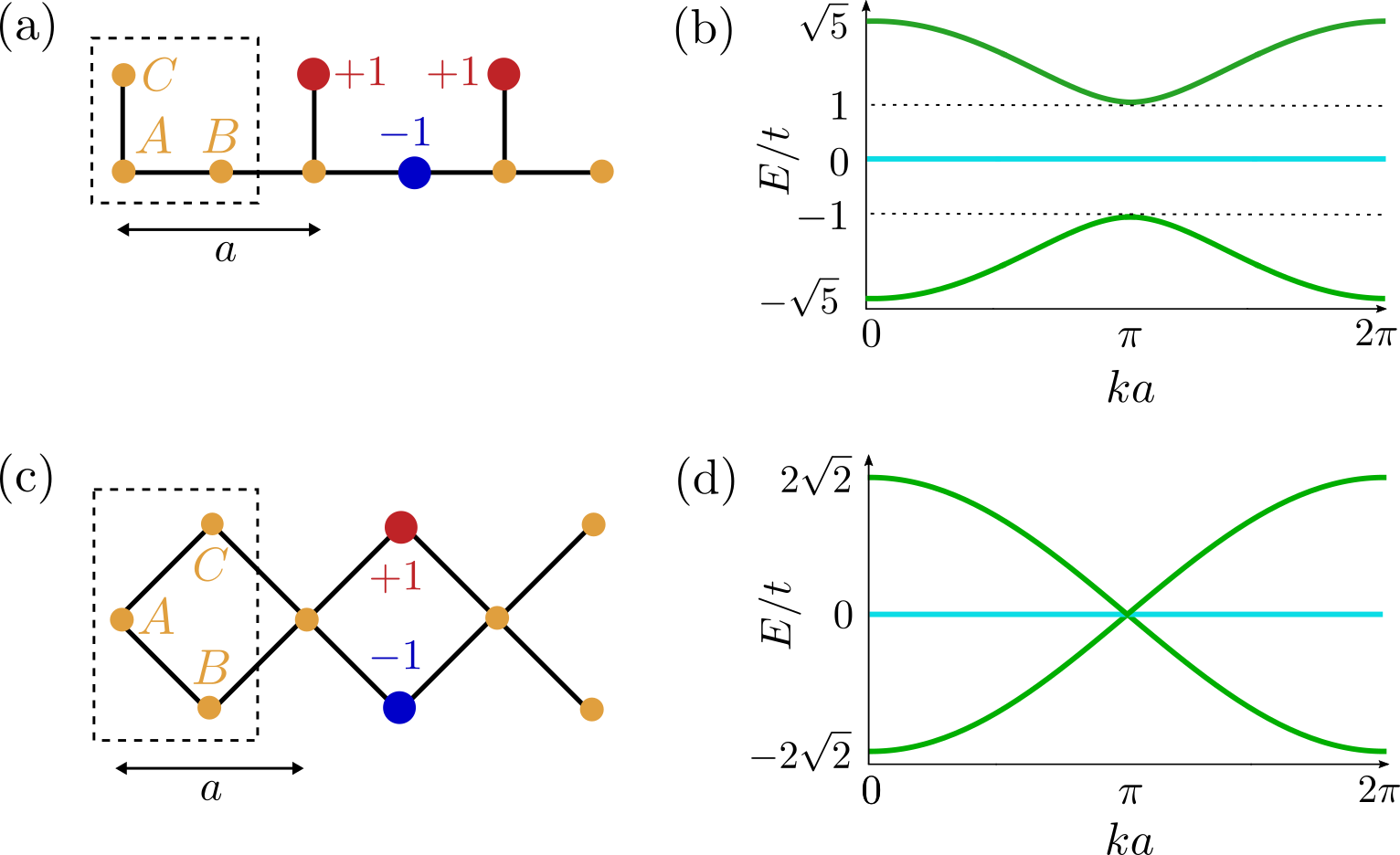}
		\caption{(a,b) Stub lattice. The unit cell (dashed rectangle) consist of three sites (orange dots) labeled $A$, $B$, and $C$.  Nearest-neighbor sites are connected by a hopping term of strength $t$ (black lines). The flat band is spanned by a set of CLSs living on three sites each (red and blue dots).  The amplitudes of the unnormalized CLSs are $+1$ ($-1$) for the red (blue) sites. The dispersive bands (green) are separated from the flat band (cyan) by an energy gap $E_{\mathrm{gap}}=t$. (c,d) Diamond lattice. Here, the CLSs have support on two sites each. The dispersive bands (green) linearly intersect the flat band (cyan).}
		\label{fig:1D_lattices}
	\end{figure}

Conventionally, the RKKY interaction is calculated in second-order perturbation theory assuming that the exchange coupling between the impurity spins and the itinerant electrons is small compared to the typical energy scale of the latter. Recently, however, systems with so-called \emph{flat bands} have attracted significant attention~\cite{Liu2014,Leykam2018}. The energy of these bands is completely independent of momentum (see Fig.~\ref{fig:1D_lattices}) or, in a weaker sense, at least approximately constant over a large range of allowed momenta. While the recent interest in flat-band systems has mainly been fueled by significant theoretical and experimental progress on Moir\'{e} materials such as twisted bilayer graphene~\cite{Bistritzer2011,Cao2018,Cao2018b,MacDonald2019,Andrei2020,Balents2020}, flat bands can also emerge as Landau levels in two-dimensional electron gases subjected to a strong magnetic field or in a variety of artificial lattice models~\cite{Sutherland1986,Lieb1989,Mielke1991,Tasaki1992,Vidal1998,Mielke1999,Vidal2000}, some of which have successfully been realized in experiments using photonic lattices or cold-atom setups~\cite{Shen2010,Apaja2010,Zhang2015,Slot2017,Xia2018,Huda2020}.
	
In the presence of flat bands, the vanishing band width and the large degeneracy make it questionable whether the conventional perturbative approach to the RKKY interaction is still applicable~\cite{note3}. This issue was first touched upon in the context of zigzag graphene nanoribbons, where exact numerical studies of edge impurities revealed unconventional features of the RKKY interaction that had not been captured by preceding analytical studies~\cite{Bunder2009,Black-Schaffer2010}. Later, Ref.~\cite{Cao2019} found unconventional first-order contributions to the RKKY interaction in partially filled graphene Landau levels via \emph{degenerate} perturbation theory. A few more recent studies calculate the standard second-order contribution to the RKKY interaction in two-dimensional flat-band lattice models (in particular, in the Lieb lattice)~\cite{Oriekhov2020,Bouzerar2021}, while Ref.~\cite{Bouzerar2022} points out that this does not capture certain flat-band effects in the Kondo-Lieb model. However, a more general understanding of RKKY effects in flat-band systems---including, in particular, insights regarding the applicability and limitations of perturbation theory---is still lacking. With this motivation, we carefully study the RKKY interaction in two simple one-dimensional (1D) flat-band systems at half filling, see Fig.~\ref{fig:1D_lattices}. We first calculate the exact RKKY interaction numerically and then compare our data to results obtained via different analytical techniques. In both our examples, we find that the RKKY interaction exhibits peculiar features that are not captured by the conventional RKKY approximation based on non-degenerate perturbation theory. Instead, we find that degenerate perturbation theory correctly reproduces our exact results if there is an energy gap between the flat and the dispersive bands, while a nonperturbative approach becomes necessary in the absence of a gap.

\section{Models}
A unit cell of the stub lattice consists of three sites labeled by $l\in\{A,B,C\}$, see Fig.~\ref{fig:1D_lattices}(a). Neighboring sites are coupled by a hopping element of strength $t>0$, such that
\begin{equation}
H_{\mathrm{stub}}=t\sum_n\left( c_{n,A}^\dagger c_{n,B}+c_{n,A}^\dagger c_{n,C}+c_{n+1,A}^\dagger c_{n,B}\right)+\mathrm{H.c.}
\label{eq:Hstub}
\end{equation}
Here, $c_{n,l}^\dagger$ ($c_{n,l}$) creates (destroys) a spinless electron on sublattice $l$ in the $n$th unit cell. Imposing periodic boundary conditions on a chain with $N$ unit cells, the Hamiltonian can be rewritten in momentum space as $H_\mathrm{stub}=\sum_k \Psi_k^\dagger\mathcal{H}(k)\Psi_k$ with $\Psi_k=\left(c_{k,A},c_{k,B},c_{k,C}\right)^T$ and
\begin{equation}
\mathcal{H}(k)=t\begin{pmatrix}0&1+e^{ika}&1\\1+e^{-ika}&0&0\\1&0&0\end{pmatrix},\label{eq:Hstub_k}
\end{equation}
where $a$ denotes the lattice spacing. The corresponding bulk spectrum consists of two dispersive bands $E_\pm(k)=\pm t\sqrt{3+2\cos{(ka)}}$ as well as one completely flat band $E_0(k)=0$ that is separated from the dispersive bands by an energy gap $E_{\mathrm{gap}}=t$, see Fig.~\ref{fig:1D_lattices}(b). The flat band is macroscopically degenerate and is spanned by a set of $N$ linearly independent states. These can be chosen to have support on only three lattice sites each: $|v_n\rangle=\left(|n,C\rangle-|n,B\rangle+|n+1,C\rangle\right)/\sqrt{3}$ for $n\in\{1,...,N\}$ and where we identify $N+1\equiv 1$ to simplify the notation. One of these so-called compact localized states (CLSs)~\cite{Sutherland1986,Leykam2018} is visualized in Fig.~\ref{fig:1D_lattices}(a). While the CLSs are chosen such that they are strictly localized, they are not mutually orthogonal. In order to construct a set of mutually orthogonal basis states for the flat band, the strict localization has to be traded in for exponential localization, e.g., by changing to a basis of maximally localized Wannier states.
	
A unit cell of the diamond lattice consists of three sites as well, see Fig.~\ref{fig:1D_lattices}(c). The Hamiltonian is given by
\begin{align}
H_{\mathrm{dia}}&=t\sum_n\Big( c_{n,A}^\dagger c_{n,B}+c_{n,A}^\dagger c_{n,C}\nonumber\\&\hspace{13.5mm} +c_{n+1,A}^\dagger c_{n,B}+c_{n+1,A}^\dagger c_{n,C}\Big)+\mathrm{H.c.}
\label{eq:Hdiamond}
\end{align}
In momentum space, this leads to $H_{\mathrm{dia}}=\sum_k \Psi_k^\dagger\mathcal{H}(k)\Psi_k$ with
\begin{equation}
\mathcal{H}(k)=t\begin{pmatrix}0&1+e^{ika}&1+e^{ika}\\1+e^{-ika}&0&0\\1+e^{-ika}&0&0\end{pmatrix}.\label{eq:Hdiamond_k}
\end{equation}
Again, the bulk spectrum consists of two dispersive bands $E_\pm(k)=\pm 2\sqrt{2}t\cos{(ka/2)}$ and a flat band $E_0(k)=0$, see Fig.~\ref{fig:1D_lattices}(d). Importantly, however, there is now no energy gap separating the flat band from the dispersive bands. Rather, the two dispersive bands linearly intersect the flat band at $ka=\pi$. The flat band can again be described in terms of a set of CLSs having support on two lattice sites each, see Fig.~\ref{fig:1D_lattices}(c). Explicitly, their wave functions are given by $|v_n\rangle=\left(|n,C\rangle-|n,B\rangle\right)/\sqrt{2}.$ Both the stub and the diamond lattice are bipartite lattices with one sublattice given by all $A$ sites and the other one by all $B$ and $C$ sites. Furthermore, we note that the flat band of the stub lattice is topologically trivial, i.e., its 1D topological invariant (winding number) is zero, while it is not meaningful to assign a topological invariant to the flat band of the diamond lattice as it is not energetically isolated.

\section{RKKY interaction}
We now consider a system of spinful electrons at zero temperature with both spin species independently described by $H_\mathrm{stub}$ or $H_\mathrm{dia}$. Throughout this work, we set the chemical potential $\mu=0$ and focus on the case of a half-filled flat band. However, we have checked that our results do not depend on the exact filling factor as long as the flat band stays partially filled. Two magnetic impurities are placed in the unit cells $n_1$ and $n_2$ at sublattice positions $\alpha$ and $\beta$, respectively. The local exchange coupling between the impurity spins and the itinerant electrons is described as $H_{\mathrm{imp}}^{(1)}+H_{\mathrm{imp}}^{(2)}$ with
\begin{align}
H_{\mathrm{imp}}^{(i)}&=\frac{\bar{J}_i}{2}\sum_{\sigma,\sigma'}\, c_{n_i,l_i,\sigma}^\dagger\, [\mathbf{S}_i \cdot\boldsymbol{\sigma}]^{\sigma\sigma'} c_{n_i,l_i,\sigma'},\label{eq:Himp}
\end{align}
where we have defined $l_1=\alpha$ and $l_2=\beta$. Compared to Eqs.~(\ref{eq:Hstub}) and (\ref{eq:Hdiamond}), the electronic creation (annihilation) operators $c_{n,l,\sigma}^\dagger$ ($c_{n,l,\sigma}$) now carry an additional spin label \mbox{$\sigma\in\{\uparrow, \downarrow\}$}. Furthermore, $\boldsymbol{\sigma}$ is the vector of Pauli matrices, $\mathbf{S}_{i}$ are classical impurity spins with $S_i=|\mathbf{S}_i|\gg 1$, and $\bar{J}_{i}\geq 0$ denotes the exchange coupling between the impurity spin and the electron spin density. For simplicity, we also define $J_i=\bar{J}_iS_i$.
	
Since there is no spin-orbit interaction in our problem, the indirect exchange interaction between the two impurity spins is isotropic and can be written as
\begin{equation}
H_{\mathrm{RKKY}}=J_{\mathrm{RKKY}}^{\alpha\beta}\hat{\mathbf{S}}_1\cdot\hat{\mathbf{S}}_2
\end{equation}
with $\hat{\mathbf{S}}_i=\mathbf{S}_i/S_i$. Here, the effective RKKY coupling constant $J_{\mathrm{RKKY}}^{\alpha\beta}\equiv J_{\mathrm{RKKY}}^{\alpha\beta}(R)$ depends on the sublattice position of the impurities and on the inter-impurity distance $R=r_2-r_1>0$ with $r_i=n_ia$. The exact RKKY coupling~\cite{note3} can be obtained from the exact ground state energies $E_{\mathrm{FM}}^{\alpha\beta}$ and $E_{\mathrm{AFM}}^{\alpha\beta}$ for the ferromagnetic (FM) and antiferromagnetic (AFM) configuration of the impurities with respect to an arbitrarily chosen spin quantization axis, say, the $z$ axis, such that $\mathbf{S}_i=(0,0,\pm S_i)$:
\begin{equation}
J_{\mathrm{RKKY}}^{\alpha\beta}=(E_{\mathrm{FM}}^{\alpha\beta}-E_{\mathrm{AFM}}^{\alpha\beta})/2.\label{eq:RKKY_energy_difference}
\end{equation}
The energies $E_{\mathrm{FM}/\mathrm{AFM}}^{\alpha\beta}$ can be computed numerically via exact diagonalization (ED)~\cite{Black-Schaffer2010} or, alternatively, via the exact lattice Green functions using the optimized algorithm presented in Appendix~\ref{app:GF_numerics}. This second approach allows us to study significantly larger system sizes while at the same time improving the numerical accuracy of our results. 
	
\begin{figure}[tb]
	\centering
	\includegraphics[width=0.85\columnwidth]{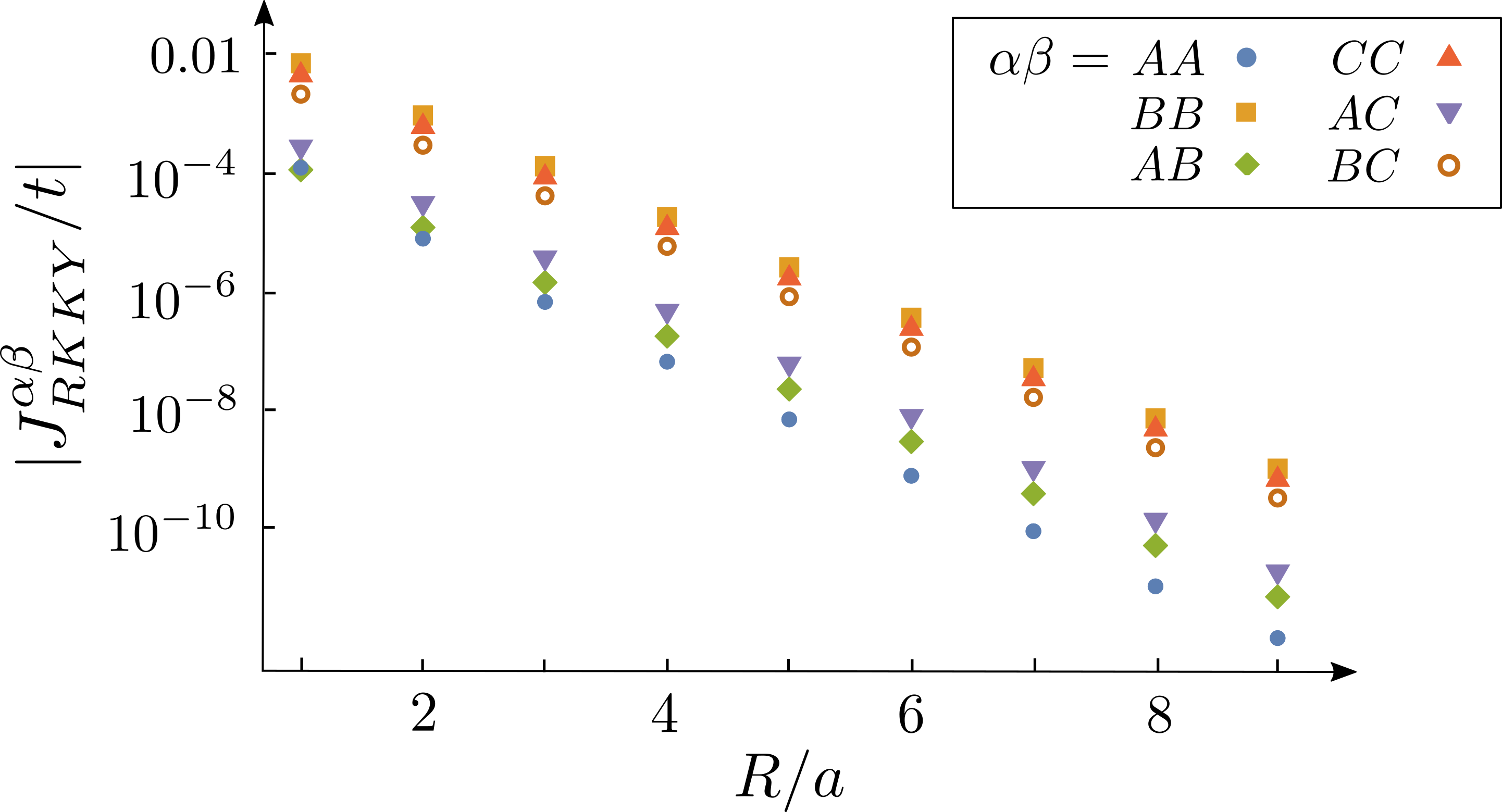}
	\caption{Absolute value of the RKKY coupling $|J_{\mathrm{RKKY}}^{\alpha\beta}|$ in the stub lattice in dependence on the inter-impurity distance $R$, calculated via ED and displayed on a logarithmic scale. For all sublattice configurations, $|J_{\mathrm{RKKY}}^{\alpha\beta}|$ decays exponentially with $R$. Here, $J_1=J_2=0.2t$.}
	\label{fig:RKKY_stub_r}
\end{figure}
	
\subsection{Stub lattice}
We start by studying the RKKY interaction in the stub lattice. Our numerical results show that $J_{\mathrm{RKKY}}^{\alpha\beta}$ decays exponentially with $R$ for all sublattice configurations, see Fig.~\ref{fig:RKKY_stub_r}. For the $AA$ configuration, this is not surprising since the flat-band states do not have support on the $A$ sublattice. As such, we expect to recover the usual Bloembergen-Rowland behavior found in conventional insulators~\cite{Bloembergen1955}. In fact, for all sublattice configurations involving at least one impurity on the $A$ sublattice, virtual transitions between the gapped dispersive bands yield the dominant contribution to the RKKY interaction. For configurations involving only the $B$ and $C$ sublattice, on the other hand, the flat-band states give an additional contribution that is responsible for the significantly larger absolute value of $J_{\mathrm{RKKY}}^{\alpha\beta}$ in these cases (see also below). However, the flat-band states are spatially localized (e.g., they can be constructed as exponentially localized Wannier states), such that their contribution is exponentially suppressed with $R$ as well. Furthermore, in accordance with the general result for bipartite lattices at half filling~\cite{Saremi2007}, we find that the ground state is FM (AFM) if the two impurities are located on the same (on different) sublattices of the bipartition.
	
\begin{figure}[bt]
	\centering
	\includegraphics[width=\columnwidth]{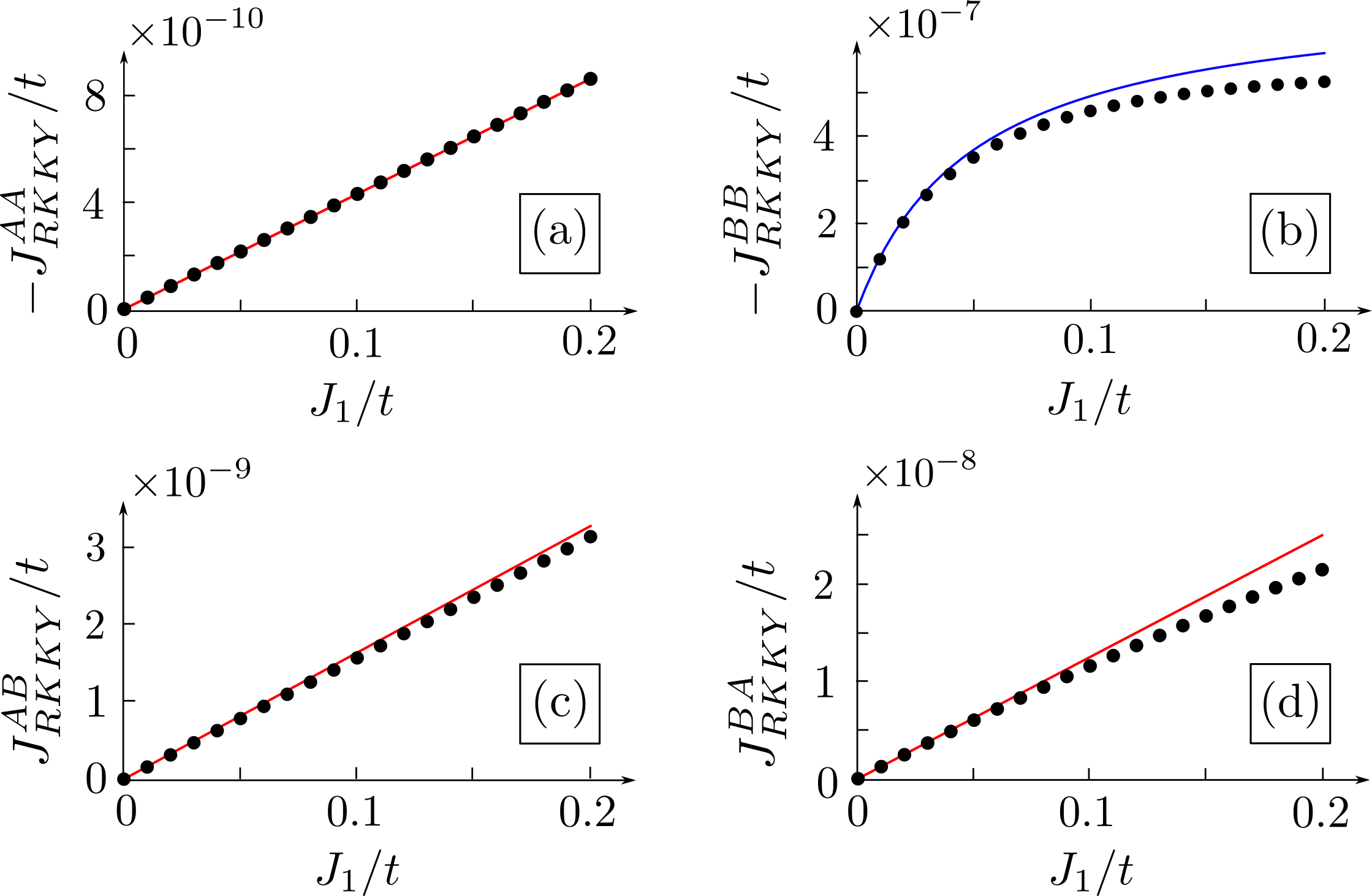}
	\caption{RKKY coupling $J_{\mathrm{RKKY}}^{\alpha\beta}$ in the stub lattice in dependence on $J_1$ calculated via ED (black) and lowest-order perturbation theory (blue: first-order, red: second-order). The standard second-order approximation [Eq.~(\ref{eq:RKKY_pert})] gives the correct lowest-order approximation for (a) $J_{\mathrm{RKKY}}^{AA}$, (c) $J_{\mathrm{RKKY}}^{AB}$, and (d) $J_{\mathrm{RKKY}}^{BA}$. The approximation in (d) is worse than in the other cases since, as $J_1$ increases, an unconventional third-order term $\propto J_1^2J_2$ originating from the flat band becomes important. (b) $J_{\mathrm{RKKY}}^{BB}$ shows an unconventional behavior due to a first-order contribution originating from the flat band, see Eq.~(\ref{eq:firstorder}). Here, $J_2/t=0.05$ and $R/a=5$~\cite{note4}.}
	\label{fig:RKKY_stub_J}
\end{figure}

To gain further insight, we study $J_{\mathrm{RKKY}}^{\alpha\beta}$ in dependence on one of the exchange coupling constants---say, $J_1$---for $J_{1,2}/t\ll 1$. 
We find that $J_{\mathrm{RKKY}}^{AA}\propto J_1$, see Fig.~\ref{fig:RKKY_stub_J}(a). This is the functional dependence expected from the standard expression for the RKKY interaction in second-order perturbation theory at zero temperature~\cite{Abrikosov1988},
\begin{equation}
J_{\mathrm{RKKY}}^{\alpha\beta}=-\frac{J_1J_2}{2\pi}\int_{-\infty}^{0}dE\,\mathrm{Im}[G_{\alpha\beta}^{(0)}(R,E)G_{\beta\alpha}^{(0)}(-R,E)],
\label{eq:RKKY_pert}
\end{equation}
where $G_{\alpha\beta}^{(0)}$ are the retarded single-particle Green functions of the unperturbed system for a single spin species~\cite{note2}. Evaluating Eq.~(\ref{eq:RKKY_pert}) by using the analytical expression for $G_{AA}^{(0)}$ (see Appendix~\ref{app:GF_stub}), we see that it reproduces the numerical result very well, see Fig.~\ref{fig:RKKY_stub_J}(a). Indeed, since the flat-band states do not have support on the $A$ sublattice, there is no reason why the standard RKKY approximation should not be valid. The flat band does not contribute to the RKKY interaction at all in this case, and the entire second-order contribution captured in Eq.~(\ref{eq:RKKY_pert}) comes from Bloembergen-Rowland transitions between the gapped dispersive bands.

In stark contrast to this, the $BB$ configuration [see Fig.~\ref{fig:RKKY_stub_J}(b)] exhibits a more complicated dependence that cannot be reproduced by Eq.~(\ref{eq:RKKY_pert}), which is in fact divergent in this case. Instead, due to the large degeneracy of the flat band, \emph{degenerate} perturbation theory (for $J_{1,2}/E_{\mathrm{gap}}=J_{1,2}/t\ll 1$) has to be used. This gives a nonvanishing first-order contribution to $J_{\mathrm{RKKY}}^{BB}$ that is responsible for the unusual $J_1$-dependence in Fig.~\ref{fig:RKKY_stub_J}(b) as well as for the significantly larger absolute value of the RKKY coupling. This first-order contribution stems entirely from intraband transitions within the flat band and is therefore only present when both impurities are located on either the $B$ or $C$ sublattice. To calculate this contribution, we apply the Gram-Schmidt orthogonalization method to the CLSs that span the flat band of the stub lattice. It is straightforward to see that we can always construct $N-2$ orthonormal basis states per spin sector that do not have support on the impurity sites, such that the entire first-order contribution is contained in an effective $2\times 2$ Hamiltonian (per spin sector) that results from projecting $H_{\mathrm{imp}}^{(1)}+H_{\mathrm{imp}}^{(2)}$ onto the remaining two basis states. For $J_{1,2}\geq 0$, we find that $J_{\mathrm{RKKY}}^{BB}$ is, to first order, given by
\begin{equation}
J_{\mathrm{RKKY}}^{BB}=-a(J_1+J_2)+\sqrt{a^2(J_1+J_2)^2-bJ_1J_2}\label{eq:firstorder}
\end{equation}
for real $R$-dependent coefficients $a,b>0$ that can be expressed as overlap integrals of the participating flat-band basis states, see Appendix~\ref{app:firstorder}. We evaluate this expression numerically and display the result in Fig.~\ref{fig:RKKY_stub_J}(b). For small  $J_{1}/t\ll 1$, we get a good agreement with the exact result. As $J_{1}/t$ gets larger, also the second-order contribution (not shown here) should be taken into account to get a better match.

Finally, we find that Eq.~(\ref{eq:RKKY_pert}) gives the correct lowest-order approximation for the $AB$ configuration ($BA$ configuration), see Fig.~\ref{fig:RKKY_stub_J}(c) [Fig.~\ref{fig:RKKY_stub_J}(d)]. As $J_{1}/t$ gets larger, an unconventional third-order contribution (not shown here) proportional to $J_1 J_2^2$ (proportional to $J_1^2 J_2$) originating from the flat band becomes important, causing visible deviations between the numerical results and the second-order approximation.
	
\begin{figure}[tb]
	\centering
	\includegraphics[width=0.85\columnwidth]{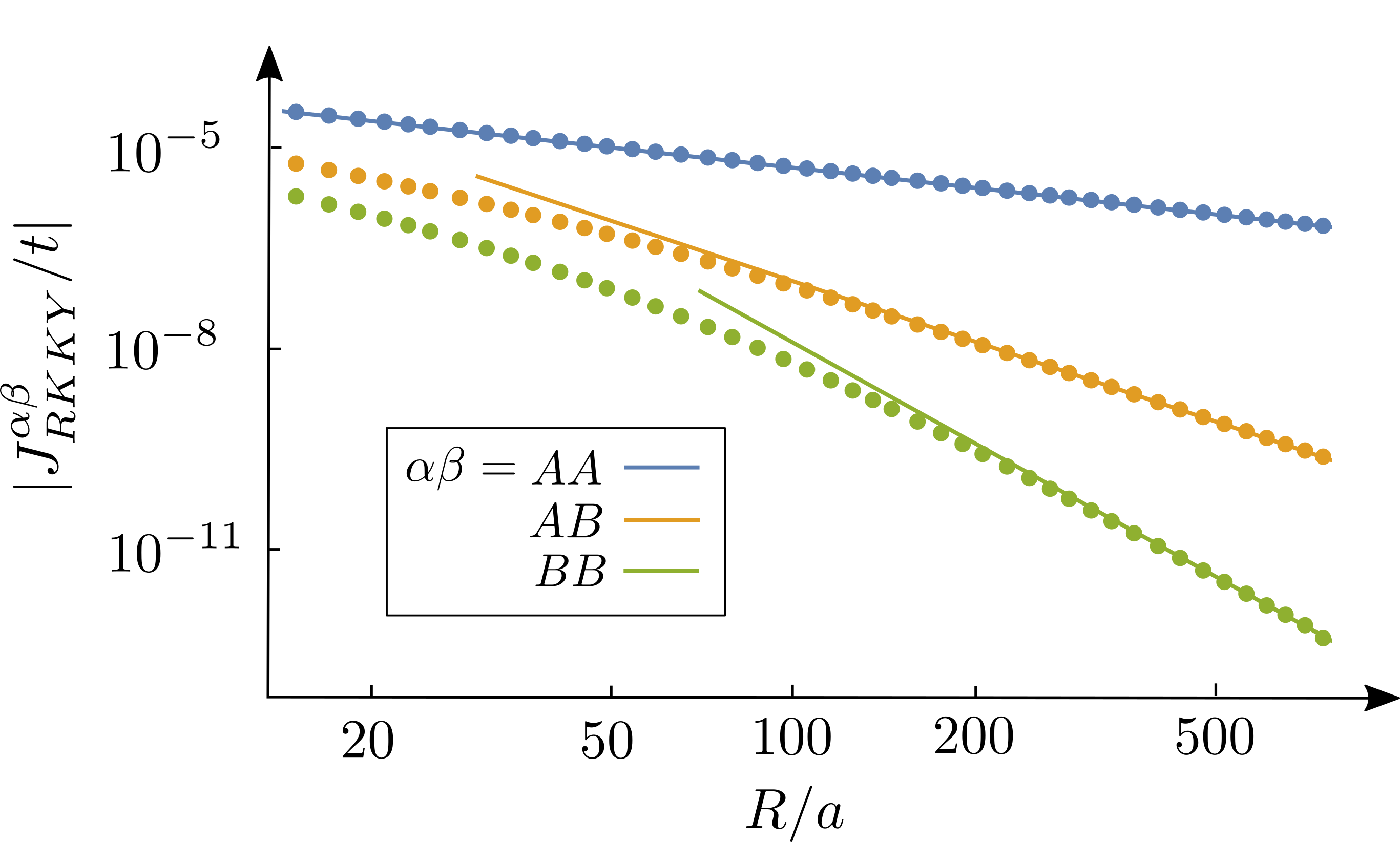}
	\caption{Absolute value of the RKKY coupling $|J_{\mathrm{RKKY}}^{\alpha\beta}|$ in the diamond lattice in dependence on the inter-impurity distance $R$ displayed in a log-log scale. The dots correspond to numerically calculated data, while the solid lines are the asymptotic analytical expressions given in Eqs.~(\ref{eq:RKKY_asymptotic_AA})--(\ref{eq:RKKY_asymptotic_BB}). For the $AA$ configuration, we find the usual $1/R$ decay that is expected in 1D metals. For the $AB$ ($BB$) configuration, the flat band leads to an unusual asymptotic $1/R^3$ ($1/R^5$) decay. Here, $J_1=J_2=0.2t$.}
	\label{fig:RKKY_diamond_r}
\end{figure}
	
\subsection{Diamond lattice}
We now proceed to study the RKKY interaction in the diamond lattice. Numerically, we find that the ground state is again FM (AFM) if the two impurities are located on the same (on different) sublattices of the bipartition. Furthermore, $J_{\mathrm{RKKY}}^{\alpha\beta}$ decays as a power law in $R$ with a leading exponent that depends on the sublattice configuration, see Fig.~\ref{fig:RKKY_diamond_r}. Since the flat-band states do not have support on the $A$ sublattice, the $AA$ configuration shows the same qualitative behavior as a conventional 1D metal, i.e., $J_{\mathrm{RKKY}}^{AA}$ decays as $1/R$. However, when one (both) impurities are placed on the $B$ or $C$ sublattices, the flat band leads to an unusual $1/R^3$ ($1/R^5$) decay. This unexpected behavior is nonperturbative in origin as discussed below. In passing, it is interesting to note that Eq.~(\ref{eq:RKKY_pert}) incorrectly predicts a $1/R$ decay for \emph{all} sublattice configurations.

\begin{figure}[tb]
	\centering
	\includegraphics[width=0.95\columnwidth]{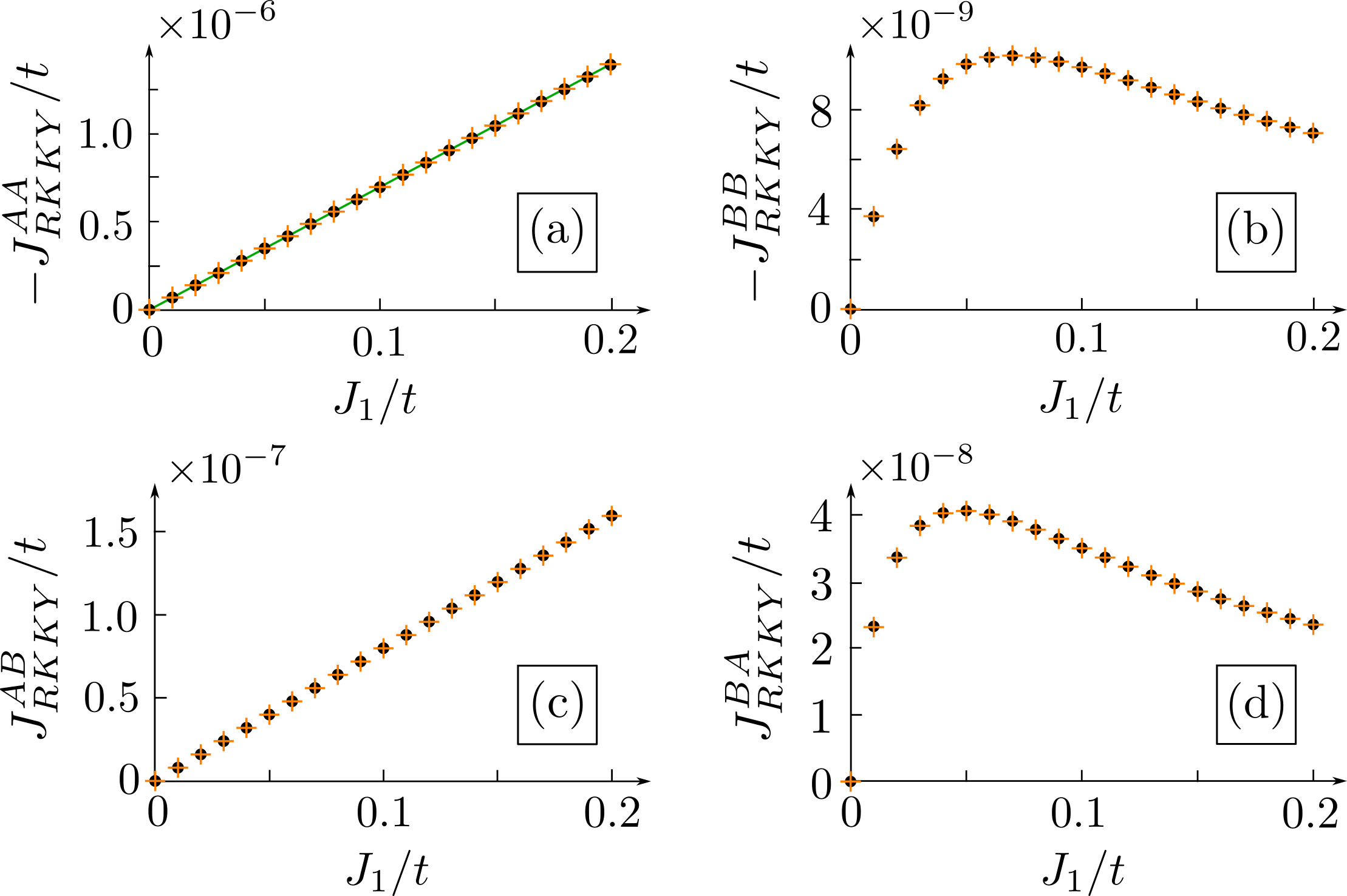}
	\caption{RKKY coupling $J_{\mathrm{RKKY}}^{\alpha\beta}$ in the diamond lattice in dependence on $J_1$ calculated via ED (black dots) and via Eq.~(\ref{eq:DeltaE_Tmatrix}) (orange crosses).  Both (a) $J_{\mathrm{RKKY}}^{AA}$ and (c) $J_{\mathrm{RKKY}}^{AB}$ are proportional to $J_1$. For (a), this is expected from Eq.~(\ref{eq:RKKY_pert}) (green line). In contrast, (b) $J_{\mathrm{RKKY}}^{BB}$ and (d) $J_{\mathrm{RKKY}}^{BA}$ increase only for very small $J_1/t$ before they start to decrease. Here, $J_2/t=0.05$ and $R/a=100$.}
	\label{fig:RKKY_diamond_J}
\end{figure}
	
Next, we can again study the dependence of $J_{\mathrm{RKKY}}^{\alpha\beta}$ on $J_1$ for small $J_{1,2}/t$. As expected from Eq.~(\ref{eq:RKKY_pert}), $J_{\mathrm{RKKY}}^{AA}\propto J_1$ [see Fig.~\ref{fig:RKKY_diamond_J}(a)]. Moreover, we find  $J_{\mathrm{RKKY}}^{AB}\propto J_1$ as well [see Fig.~\ref{fig:RKKY_diamond_J}(c)], but Eq.~(\ref{eq:RKKY_pert}) fails to predict the corresponding slope. The $BB$ and $BA$ configurations [see Figs.~\ref{fig:RKKY_diamond_J}(b) and (d)] show an even more peculiar behavior: Here, the RKKY interaction first grows rapidly for very small $J_1/t$ but then decreases. To understand this unexpected behavior, we use the $T$-matrix formalism to obtain an exact expression for the impurity-induced shift of the ground state energy. Since spin is conserved, we can treat the two spin sectors individually in the following discussion. In Appendix~\ref{app:Dos}, we show that the impurity-induced change in the density of states for a fixed spin sector and impurity configuration can be written as
\begin{equation}
\Delta\rho_{c,\sigma}^{\alpha\beta}(E)=-\frac{1}{\pi}\mathrm{Im}\,\mathrm{tr}\sum_{i,j}P(r_j-r_i,E)T_{ij}^{\alpha\beta,c,\sigma}(E)\label{eq:dos}
\end{equation}
for $c\in\{\mathrm{FM},\mathrm{AFM}\}$, $P(r,E)=\int\frac{dk}{2\pi}[G^{(0)}(k,E)]^2 e^{ikr}$, and where $G^{(0)}(k,E)$ is the retarded momentum-space Green function for a single spin sector of the unperturbed system~\cite{note2}. Furthermore, $T_{ij}^{\alpha\beta,c,\sigma}(E)$ for $i,j\in\{1,2\}$ are the components of the standard two-impurity $T$-matrix~\cite{Economou2006} that contains all information about the impurities.  Through the $T$-matrix, $\Delta\rho_{c,\sigma}^{\alpha\beta}(E)$ implicitly depends on the inter-impurity distance $R$ and can be used to express the exact $J_{\mathrm{RKKY}}^{\alpha\beta}$ as
\begin{equation}
J_{\mathrm{RKKY}}^{\alpha\beta}=\frac{1}{2}\sum_\sigma\int_{-\infty}^0 dE\,E\, [\Delta\rho_{\mathrm{FM},\sigma}^{\alpha\beta}(E)-\Delta\rho_{\mathrm{AFM},\sigma}^{\alpha\beta}(E)]\label{eq:DeltaE_Tmatrix}.
\end{equation}
Performing the integration over energy numerically, we find that this reproduces the results obtained by ED very well, see Fig.~\ref{fig:RKKY_diamond_J}. Even more interestingly, Eq.~(\ref{eq:DeltaE_Tmatrix}) can be treated analytically in the limit of large $R/a$ and small but finite $J_{1,2}/t$. After some straightforward calculations outlined in Appendix~\ref{app:RKKYasymptotic}, we find
\begin{align}
J_{\mathrm{RKKY}}^{AA}&\approx-\frac{J_1J_2}{16\sqrt{2}t\pi (R/a)}\quad\mathrm{for}\ \  R/a\gg 1,\label{eq:RKKY_asymptotic_AA}\\
J_{\mathrm{RKKY}}^{AB}&\approx\frac{J_1t}{2\sqrt{2}\pi J_2 (R/a)^3}\,\quad\mathrm{for}\ \  R/a\gg t/J_2,\label{eq:RKKY_asymptotic_AB}\\
J_{\mathrm{RKKY}}^{BB}&\approx-\frac{12\sqrt{2}\,t^3}{\pi J_1 J_2 (R/a)^5}\ \quad\mathrm{for}\ \ R/a\gg t/J_{1,2}.\label{eq:RKKY_asymptotic_BB}
\end{align}
These expressions nicely approach our numerical data for large $R/a$, see Fig.~\ref{fig:RKKY_diamond_r}. Importantly, for sufficiently large $R/a$, Eqs.~(\ref{eq:RKKY_asymptotic_AA})--(\ref{eq:RKKY_asymptotic_BB}) hold down to arbitrarily small but finite $J_{1,2}$. In this sense, the results for the $AB$ and $BB$ configuration are highly nonperturbative. More generally, all configurations with at least one impurity on the $B$ or $C$ sublattice exhibit such a nonperturbative behavior. On the other hand, it should be stressed that, for any finite $R/a$, $J_{\mathrm{RKKY}}^{\alpha\beta}$ remains well-behaved and goes to zero as $J_{1,2}\rightarrow 0$ for all sublattice configurations. This behavior can for example be observed in Figs.~\ref{fig:RKKY_diamond_J}(b) and (d), where $J_{\mathrm{RKKY}}^{\alpha\beta}$ changes its trend with respect to $J_1$ as $t/J_{1}$ becomes of the order of $R/a$.

\section{Conclusions}
We have studied the RKKY interaction in two 1D flat-band models at half filling. In our first example---the stub lattice---we have found an unconventional first-order contribution to the RKKY interaction due to the degeneracy of the isolated flat band. In our second example---the diamond lattice---the absence of an energy gap between the flat and the dispersive bands leads to a breakdown of perturbation theory altogether, and nonperturbative contributions cause the RKKY interaction to decay more rapidly with the inter-impurity distance than na\"{i}vely expected. Our results illustrate that the RKKY interaction in flat-band systems can exhibit unexpected features and has to be treated with care. While we have focused on 1D toy models for analytical and numerical simplicity, both the scenario of an isolated flat band and of intersecting flat and Dirac-like bands can also occur in experimentally relevant two-dimensional materials such as twisted bilayer~\cite{Bistritzer2011,Cao2018,Cao2018b,MacDonald2019,Andrei2020,Balents2020} and trilayer~\cite{Khalaf2019,Carr2020,Lei2021,Kim2021,Shen2022,Li2022} graphene, respectively. It would be interesting to extend our calculations to flat-band systems in two dimensions, and, especially, to topologically nontrivial flat bands due to the close relation between Wannier state localization and band topology~\cite{Bergholtz2013}.

It is furthermore interesting to explore how our results are modified by, e.g., electron-electron interactions or disorder. We leave these questions to future work. Moreover, we have focused on the case of a perfectly flat band at half filling, where we expect flat-band effects to be the most pronounced. Nevertheless, it would also be interesting to study more general fillings and small deviations from perfect flatness. In general, as long as the exchange coupling constants are larger than any additional energy scale resulting, e.g., from a small but finite bandwidth, we expect the unconventional effects reported here to persist.
	
\textit{Acknowledgments.} %
We thank Henry F. Legg and Martin Claassen for helpful discussions. This work was supported by the Deutsche Forschungsgemeinschaft via RTG 1995, the Swiss National Science Foundation (SNSF) and NCCR QSIT, and by the Deutsche Forschungsgemeinschaft (DFG, German Research Foundation) under Germany's Excellence Strategy - Cluster of Excellence Matter and Light for Quantum Computing (ML4Q) EXC 2004/1 - 390534769. We acknowledge support from the Max Planck-New York City Center for Non-Equilibrium Quantum Phenomena. This project received funding from the European Union’s Horizon 2020 research and innovation program (ERC Starting Grant, grant agreement No 757725). Simulations were performed with computing resources granted by RWTH Aachen University under projects rwth0752 and rwth0841.

\pagebreak
\begin{widetext}
\appendix

\section{Matsubara Green functions for the stub lattice}
\label{app:GF_stub}

In this appendix, we list the Matsubara Green functions for the stub lattice. To simplify the notation, we set $a=t=1$ throughout all appendices unless specified otherwise. The Matsubara Green functions are defined as $G^{(0)}(k,i\omega)=[i\omega-\mathcal{H}(k)]^{-1}$, where $\mathcal{H}(k)$ is given in Eq.~(\ref{eq:Hstub_k}). Evaluating this formula leads us directly to
\begin{align}
G^{(0)}(k, i\omega) &= \frac{1}{2[3+2\cos(k)]} \left(\begin{array}{ccc}
-\frac{2 [3+2\cos(k)] i\omega}{\omega^2+3+2\cos(k)} & -\frac{2(1+e^{ik})[3+2\cos(k)]}{\omega^2+3+2\cos(k)} & -\frac{2[3+2\cos(k)]}{\omega^2+3+2\cos(k)} \\
-\frac{2[3+2\cos(k)] (1+e^{-ik})}{\omega^2+3+2\cos(k)} &
-\frac{4i\omega[1+\cos(k)]}{\omega^2+3+2\cos(k)}+\frac{2}{i\omega} &
-\frac{2i\omega (1+e^{-ik})}{\omega^2+3+2\cos(k)}-\frac{4[1+\cos(k)]}{i\omega (e^{ik}+1)} \\
-\frac{2[3+2\cos(k)]}{\omega^2+3+2\cos(k)} & -\frac{2i\omega (1+e^{ik})}{\omega^2+3+2\cos(k)}-\frac{4(1+\cos(k)]}{i\omega(1+e^{-ik})} &
\frac{-2i\omega}{\omega^2+3+2\cos(k)}+\frac{4[1+\cos(k)]}{i\omega}
\end{array}\right)\,
\label{eq:Green_stub}
\end{align}
in the basis $\Psi_k=\left(c_{k,A},c_{k,B},c_{k,C}\right)^T$. From this, we can obtain the real-space Matsubara Green functions $G_{\alpha\beta}^{(0)}(r,i\omega)=[G^{(0)}(r,i\omega)]_{\alpha\beta}$ by a Fourier transformation
\begin{align}
G^{(0)}_{\alpha \beta}(r, i\omega) = \int_{-\pi}^\pi \frac{dk}{2\pi} \, G^{(0)}_{\alpha \beta}(k, i\omega) \, e^{ikr} .
\end{align}
For $r\geq0$ and $\omega\neq0$, we find: 
\allowdisplaybreaks
\begin{align}
&G_{AA}^{(0)}(r,i\omega) = - \frac{i  \,  2^{-r} \omega (-\omega^2-3+\eta)^r}{\eta},\\
&G_{AB}^{(0)}(r,i\omega) = \frac{2^{-r-1}\left(-\omega^2-5+\eta\right) \left(-\omega^2-3+\eta\right)^r}{\omega^2+5},\\
&G_{BA}^{(0)}(r,i\omega) = \frac{2^{-r-1} \left(\omega^2+5+\eta\right) \left(-\omega^2-3+\eta\right)^r}{\omega^2+5}-\delta_{r,0} \, ,
\end{align}
where we have defined $\eta= \sqrt{\left(\omega^2+1\right)\left(\omega^2+5\right)}$. The missing components are given by $G_{CA}^{(0)}=G_{AC}^{(0)}=\frac{1}{i\omega} G_{AA}^{(0)}$, $G_{BB}^{(0)}=(1+\frac{1}{\omega^2}) G_{AA}^{(0)}$, $G_{BC}^{(0)}=\frac{1}{i\omega} G_{BA}^{(0)}$, $G_{CB}^{(0)}=\frac{1}{i\omega} G_{AB}^{(0)}$ and $G_{CC}^{(0)}=\frac{-1}{\omega^2} G_{AA}^{(0)}+\frac{\delta_{r,0}}{i\omega}$. The Green functions for $r<0$ can be found from the relation $G_{\alpha\beta}^{(0)}(r,i\omega)=[G_{\beta\alpha}^{(0)}(-r,-i\omega)]^*$. \\

These Green functions were used to evaluate Eq.~(\ref{eq:RKKY_pert}) in Matsubara space using
\begin{align}
J_{\mathrm{RKKY}}^{\alpha\beta}&=-\frac{J_1J_2}{2\pi}\int_{-\infty}^{0}dE\,\mathrm{Im}[G_{\alpha\beta}^{(0)}(R,E)G_{\beta\alpha}^{(0)}(-R,E)] \nonumber \\
&=\frac{J_1 J_2}{2\pi} \int_{0}^\infty d\omega \, G^{(0)}_{\alpha\beta}(R, i\omega) G^{(0)}_{\beta\alpha}(-R, i\omega) \, .\label{eq:rkky_pert_matsubara}
\end{align}
Here, we exploited that the integrand of the latter integral is always real in our case.

\section{Degenerate perturbation theory for the stub lattice}
\label{app:firstorder}

In this appendix, we derive Eq.~(\ref{eq:firstorder}). Since the spin along the $z$ direction is conserved in our problem, we consider the two spin sectors independently in the following. In a system of $N$ lattice sites with periodic boundary conditions, the flat band (in a given spin sector) is spanned by $N$ states of the form
\begin{equation}
|v_n\rangle=\left(|n,C\rangle-|n,B\rangle+|n+1,C\rangle\right)/\sqrt{3}
\end{equation}
for $n\in\{1,...,N\}$ and where we identify $N+1\equiv 1$ to simplify the notation. Let us start by considering the case where both impurities are placed on the $B$ sublattice. Like in the main text, let us label the unit cells where the impurities are placed by $n_1$, $n_2$. It is then straightforward to see that only the two states $|v_{n_1}\rangle$, $|v_{n_2}\rangle$ have support on the impurity sites. We now order all flat-band states such that $|v_{n_1}\rangle$, $|v_{n_2}\rangle$ are the last states of the set and apply the Gram-Schmidt orthonormalization procedure in order to obtain an orthonormal basis for the flat band. Clearly, the first $N-2$ states of the basis we obtain will have zero support on the impurity sites. Only the last two states of the orthonormal set, let us call them $|\tilde{v}_{n_1}\rangle$ and $|\tilde{v}_{n_2}\rangle$, have support on the impurity sites. Therefore, we can now obtain an effective Hamiltonian by projecting the impurity Hamiltonian given in Eq.~(\ref{eq:Himp}) of the main text onto $|\tilde{v}_{n_1}\rangle$ and $|\tilde{v}_{n_2}\rangle$. For the FM configuration, we obtain
\begin{equation}
\mathcal{H}_{\mathrm{eff,FM,\sigma}}=\frac{\sigma}{2}\begin{pmatrix}
J_1\langle\tilde{v}_{n_1}|n_1,B\rangle\langle n_1,B|\tilde{v}_{n_1}\rangle & J_1\langle\tilde{v}_{n_1}|n_1,B\rangle\langle n_1,B|\tilde{v}_{n_2}\rangle \\
J_1\langle\tilde{v}_{n_2}|n_1,B\rangle\langle n_1,B|\tilde{v}_{n_1}\rangle &
J_1\langle\tilde{v}_{n_2}|n_1,B\rangle\langle n_1,B|\tilde{v}_{n_2}\rangle+J_2\langle\tilde{v}_{n_2}|n_2,B\rangle\langle n_2,B|\tilde{v}_{n_2}\rangle
\end{pmatrix}.
\end{equation}
Let us introduce the shorthand notations $w=\langle\tilde{v}_{n_1}|n_1,B\rangle$, $y=\langle\tilde{v}_{n_2}|n_1,B\rangle$, $z=\langle\tilde{v}_{n_2}|n_2,B\rangle$. These coefficients are real since the original $|v_n\rangle$ were real. Furthermore, they satisfy the relation $w^2+y^2=z^2$. The eigenvalues of $\mathcal{H}_{\mathrm{eff,FM,\sigma}}$ are then given by
\begin{equation}
E_{\pm,\mathrm{FM},\sigma}=\frac{\sigma}{4}\left[z^2(J_1+J_2)\pm |z| \sqrt{z^2(J_1-J_2)^2+4y^2J_1J_2}\right].
\end{equation}
Using $y^2\leq z^2$ and assuming that $J_{1,2}\geq 0$, we find that $E_{\pm,\mathrm{FM},1}>0$ and $E_{\pm,\mathrm{FM},-1}<0$. The eigenvalues for the AFM configuration are obtained by sending $J_2\rightarrow -J_2$:
\begin{equation}
E_{\pm,\mathrm{AFM},\sigma}=\frac{\sigma}{4}\left[z^2(J_1-J_2)\pm |z| \sqrt{z^2(J_1+J_2)^2-4y^2J_1J_2}\right].
\end{equation}
In this case, we find that $E_{+,\mathrm{AFM},1},E_{-,\mathrm{AFM},-1}>0$ and $E_{+,\mathrm{AFM},-1},E_{-,\mathrm{AFM},1}<0$.

To calculate the RKKY coupling within this approximation, we now subtract the sum of all negative eigenvalues in the AFM configuration from the sum of all negative eigenvalues in the FM configuration:
\begin{align}
J_{\mathrm{RKKY}}^{BB}&=\frac{1}{2}\left[E_{+,\mathrm{FM},-1}+E_{-,\mathrm{FM},-1}-\left(E_{+,\mathrm{AFM},-1}+E_{-,\mathrm{AFM},1}\right)\right]\nonumber\\&=-\frac{z^2}{4}(J_1+J_2)+\frac{1}{4}\sqrt{-4y^2z^2J_1J_2+z^4(J_1+J_2)^2}.
\end{align}
Identifying $a=z^2/4$ and $b=y^2z^2/4$, we recover Eq.~(\ref{eq:firstorder}). Note that the square root is always real since $y^2\leq z^2$.

For completeness, we also briefly discuss impurities on the $C$ sublattice. If an impurity is located on the $C$ sublattice in the unit cell $n_1$, both the state $|v_{n_1}\rangle$ and the state $|v_{n_1-1}\rangle$ have support on the impurity site. Therefore, the argument presented above cannot be carried over directly. However, we can simply work in terms of even and odd superpositions $\left(|v_{n_1-1}\rangle\pm |v_{n_1}\rangle\right)/\sqrt{2}$, where now only the even superposition has support on the impurity site. Now we can repeat similar steps as presented above in order to calculate the RKKY coupling.

\section{Matsubara Green functions for the diamond lattice}
\label{app:GF}

In this appendix, we give the Matsubara Green functions for the diamond lattice. The Matsubara Green functions in momentum space are defined via $G^{(0)}(k,i\omega)=[i\omega-\mathcal{H}(k)]^{-1}$, where $\mathcal{H}(k)$ is given in Eq.~(\ref{eq:Hdiamond_k}). This leads us to
\begin{align}
G^{(0)}(k, i\omega) = \frac{1}{4} \left(\begin{array}{ccc}
\frac{-4i\omega}{\omega^2+8\cos^2(k/2)}& \frac{-8\cos(k/2) e^{ik/2}}{\omega^2 + 8 \cos^2(k/2)} & \frac{-8\cos(k/2) e^{ik/2}}{\omega^2 + 8 \cos^2(k/2)}\\
\frac{-8\cos(k/2) e^{-ik/2}}{\omega^2 + 8 \cos^2(k/2)} & \frac{2}{i\omega} - \frac{2i\omega}{\omega^2 + 8 \cos^2(k/2)}& -\frac{2}{i\omega} - \frac{2i\omega}{\omega^2 + 8 \cos^2(k/2)}\\
\frac{-8\cos(k/2) e^{-ik/2}}{\omega^2 + 8 \cos^2(k/2)} & -\frac{2}{i\omega} - \frac{2i\omega}{\omega^2 + 8 \cos^2(k/2)}& \frac{2}{i\omega} - \frac{2i\omega}{\omega^2 + 8 \cos^2(k/2)}
\end{array}
\right),
\end{align}
where we used the same basis as in Appendix~\ref{app:GF_stub}. Performing a Fourier transform in the same way as in Appendix~\ref{app:GF_stub}, we find for $r\geq 0$ and $\omega \neq 0$:
\begin{align}
&G_{AA}^{(0)}(r,i\omega)=\frac{-i\left[-1+\frac{\omega}{4}\left(-\omega+\gamma\right)\right]^r}{\gamma},\\
&G_{AB}^{(0)}(r,i\omega)=\frac{4^{-1-r}(\omega-\gamma)\left[-4+\omega\left(-\omega+\gamma\right)\right]^r}{\gamma},\\
&G_{BA}^{(0)}(r,i\omega)=\frac{4^{-1-r}(\omega+\gamma)\left[-4+\omega\left(-\omega+\gamma\right)\right]^r}{\gamma}-\frac{\delta_{r,0}}{2},
\end{align}
where we have defined $\gamma=\mathrm{sgn}(\omega)\sqrt{8+\omega^2}$. The other components can be obtained from the above as $G_{AC}^{(0)}=G_{AB}^{(0)}$, $G_{CA}^{(0)}=G_{BA}^{(0)}$, $G_{BB}^{(0)}= G_{CC}^{(0)} =\frac{1}{2} (G_{AA}^{(0)} + \frac{\delta_{r,0}}{i\omega})$, and $G_{BC}^{(0)}=G_{CB}^{(0)} = \frac{1}{2} (G_{AA}^{(0)} - \frac{\delta_{r,0}}{i\omega})$. The Green functions for $r<0$ can again be found from the relation $G_{\alpha\beta}^{(0)}(r,i\omega)=[G_{\beta\alpha}^{(0)}(-r,-i\omega)]^*$. In the main text, we use these Green functions to evaluate Eqs.~(\ref{eq:RKKY_pert}) and (\ref{eq:DeltaE_Tmatrix}).

\section{Impurity-induced change in the density of states}
\label{app:Dos}

In this appendix, we derive Eq.~(\ref{eq:dos}). For simplicity, we use the short-hand notation $T_{ij}^{\alpha\beta,c,\sigma}(E)\equiv T_{ij}(E)$ throughout this appendix, keeping in mind that the two-impurity $T$-matrix  depends on the spin sector $\sigma\in\{\uparrow,\downarrow\}$ we consider, on the relative orientation $c\in\{\mathrm{FM},\mathrm{AFM}\}$ and the sublattice positions $\alpha,\beta$ of the impurities, as well as, implicitly, on the inter-impurity distance $R$. The full Green function $G^{\alpha\beta,c,\sigma}(r,r',E)\equiv G(r,r',E)$ of the perturbed system (for a single spin sector and a fixed impurity configuration) can then be expressed as
\begin{equation}
G(r,r',E)=G^{(0)}(r-r',E)+\sum_{i,j}G^{(0)}(r-r_i,E)T_{ij}(E)G^{(0)}(r_j-r',E),\label{eq:gf_exact}
\end{equation}
where $i,j\in\{1,2\}$ and $G^{(0)}(r,E)$ is the retarded Green function for a single spin sector of the unperturbed system. We note that, in our model, the unperturbed Hamiltonian and therefore also the unperturbed Green functions do not depend on spin, such that the spin-dependence enters solely through the $T$-matrix. The four components of the two-impurity $T$-matrix are given by~\cite{Economou2006}:
	\begin{align}
	T_{11}(E)&=\left[\mathbb{1}-V_1G^{(0)}(0,E)-V_1G^{(0)}(-R,E)T_2^{(0)}(E)G^{(0)}(R,E)\right]^{-1}V_1,\\
	T_{12}(E)&=T_1^{(0)}(E)G^{(0)}(-R,E)T_{22}(E),\\
	T_{21}(E)&=T_2^{(0)}(E)G^{(0)}(R,E)T_{11}(E),\\
	T_{22}(E)&=\left[\mathbb{1}-V_2G^{(0)}(0,E)-V_2G^{(0)}(R,E)T_1^{(0)}(E)G^{(0)}(-R,E)\right]^{-1}V_2.
	\end{align}
	Here, we have defined $V_1=\frac{\sigma J_1}{2}|\alpha\rangle\langle\alpha|$ and $V_2= \frac{c\sigma J_2}{2}|\beta\rangle\langle\beta|$, where $|l\rangle\langle l|$ is a projector on the sublattice $l\in\{A,B,C\}$ and where we define $\sigma\in\{\uparrow,\downarrow\}\equiv\{+1,-1\}$ and $c\in\{\mathrm{FM},\mathrm{AFM}\}\equiv\{+1,-1\}$. Moreover, we have defined the single-impurity $T$-matrices as 
	\begin{equation}
	T_i^{(0)}=\left[\mathbb{1}-V_i G^{(0)}(0,E)\right]^{-1}V_i.
	\end{equation}
	
	The exact Green function given in Eq.~(\ref{eq:gf_exact}) can then be used to obtain the impurity-induced change in the local density of states $\Delta\rho_{c,\sigma}^{\alpha\beta}(r,E)\equiv \Delta\rho(r,E)$ for a single spin sector and a fixed impurity configuration:
	\begin{align}
	\Delta\rho(r,E)&=-\frac{1}{\pi}\mathrm{Im}\,\mathrm{tr}[G(r,r,E)-G^{(0)}(0,E)]\nonumber\\&=-\frac{1}{\pi}\mathrm{Im}\,\mathrm{tr}\Big[\sum_{i,j}G^{(0)}(r-r_i,E)T_{ij}(E)G^{(0)}(r_j-r,E)\Big].
	\end{align}
	From this, we obtain the total change of the density of states $\Delta\rho_{c,\sigma}^{\alpha\beta}(E)\equiv \Delta\rho(E)$ for a single spin sector and a fixed impurity configuration as
	\begin{align}
	\Delta\rho(E)&=\int dr\, \Delta\rho(r,E)\nonumber\\&=-\frac{1}{\pi}\mathrm{Im}\,\mathrm{tr} \sum_{i,j}\int dr\,G^{(0)}(r-r_i,E)T_{ij}(E)G^{(0)}(r_j-r,E)\nonumber\\&=-\frac{1}{\pi}\mathrm{Im}\,\mathrm{tr} \sum_{i,j}\int \frac{dk}{2\pi}\int \frac{dk'}{2\pi}\int dr\,e^{i(k'-k)r}G^{(0)}(k,E)G^{(0)}(k',E)e^{ikr_j}e^{-ik'r_i}T_{ij}(E)\nonumber\\&=-\frac{1}{\pi}\mathrm{Im}\,\mathrm{tr}\sum_{i,j}\int \frac{dk}{2\pi}[G^{(0)}(k,E)]^2 e^{ik(r_j-r_i)}T_{ij}(E).
	\end{align}
	After reinstating the indices, this gives Eq.~(\ref{eq:dos}) of the main text.

	\section{Asymptotic expressions for the RKKY coupling}
	\label{app:RKKYasymptotic}
	
	In this appendix, we extract the asymptotic behavior of Eq.~(\ref{eq:DeltaE_Tmatrix}) in the limit of large $R$. For convenience, we write $J_{\mathrm{RKKY}}^{\alpha\beta}=\sum_{i,j}J_{\mathrm{RKKY}}^{\alpha\beta,ij}$ with
	\begin{align}
	&J_{\mathrm{RKKY}}^{\alpha\beta,ij}=\frac{1}{2}\sum_\sigma\int_{-\infty}^0 dE\,E\, [\Delta\rho_{\mathrm{FM},\sigma}^{\alpha\beta,ij}(E)-\Delta\rho_{\mathrm{AFM},\sigma}^{\alpha\beta,ij}(E)], \label{eq:integral}\\
	&\Delta\rho^{\alpha\beta,ij}_{c,\sigma}(E)=-\frac{1}{\pi}\mathrm{Im}\,\mathrm{tr}\int \frac{dk}{2\pi}[G^{(0)}(k,E)]^2 e^{ik(r_j-r_i)}T_{ij}^{\alpha\beta,c,\sigma}(E).
	\end{align}
	In the following we will solve the integral in Eq.~(\ref{eq:integral}) in Matsubara space using
	\begin{align}
	\int_{-\infty}^0 dE\,E\, \Delta\rho^{\alpha\beta,ij}_{c,\sigma}(E) &= \int_{-\infty}^\infty dE\,E\, \Delta\rho^{\alpha\beta,ij}_{c,\sigma}(E) f(E) \nonumber \\
	&= -\frac{1}{\pi}\mathrm{Im}\,\mathrm{tr}\int_{-\infty}^\infty dE\,E\,   \int  \frac{dk}{2\pi}[G^{(0)}(k,E)]^2 e^{ik(r_j-r_i)}T_{ij}^{\alpha\beta,c,\sigma}(E) f(E) \nonumber \\
	&= \frac{1}{\pi}\mathrm{Im}\,\mathrm{tr} \frac{2\pi i }{\beta} \sum_{i\omega} i\omega\,   \int  \frac{dk}{2\pi}[G^{(0)}(k,i\omega)]^2 e^{ik(r_j-r_i)}T_{ij}^{\alpha\beta,c,\sigma}(i\omega)  \, e^{i \omega 0^+} \nonumber \\
	&= -\frac{1}{\pi}\mathrm{Im}\,\mathrm{tr}  \int_0^\infty d\omega \, \omega\,   \int  \frac{dk}{2\pi}[G^{(0)}(k,i\omega)]^2 e^{ik(r_j-r_i)}T_{ij}^{\alpha\beta,c,\sigma}(i\omega) \, e^{i \omega 0^+} \, ,\label{eq:rkky_exact_matsubara}
	\end{align}
	where the last step exploits that we assume the system to be at zero temperature and where $f(E)$ denotes the Fermi-Dirac distribution function. The unperturbed Green functions for the diamond lattice that enter the above expression are given in Appendix~\ref{app:GF}.
	
	We start by discussing the $AA$ configuration. In this case, the flat band is not affected by the impurities and we can safely expand the full $T$-matrix in orders of $J_{1,2}$. Since the unperturbed system is time-reversal symmetric, first-order contributions to the RKKY coupling cancel when the two spin sectors are added up. As such, to lowest order, the RKKY coupling is given by second-order terms $\propto J_1J_2$. These terms are contained within the off-diagonal contributions 
	\begin{equation}
	J_{\mathrm{RKKY}}^{AA,12}=J_{\mathrm{RKKY}}^{AA,21}=-\frac{J_1J_2}{2\pi}\int_0^\infty d\omega\,(-i\omega)\, G^{(0)}_{AA}(R,i\omega)P^{(0)}_{AA}(-R,i\omega),\label{eq:sm_JAA}
	\end{equation}
	where we have introduced the short-hand notation $P^{(0)}(r,i\omega)=\int \frac{dk}{2\pi}[G^{(0)}(k,i\omega)]^2 e^{ikr}$ and where we have already used that $G^{(0)}_{AA}$ ($P^{(0)}_{AA}$) is purely imaginary (real). Plugging in the Green functions given in Appendix~\ref{app:GF}, we find that  Eq.~(\ref{eq:sm_JAA}) gives us
	\begin{equation}
	J_{\mathrm{RKKY}}^{AA}=-\frac{J_1J_2}{\pi}\int_0^\infty d\omega\, \omega \left[\frac{16^{-R}(4+\omega^2-\omega\gamma)^{2R}(\omega+2\gamma R)}{\gamma^4}\right]\label{eq:EAA_Tmatrix}
	\end{equation}
	with $\gamma=\sqrt{8+\omega^2}$. The integral converges on a scale $\propto 1/R$, which is why we change the integration variable to $\omega'=\omega R$ and then cut the upper integration limit at some finite constant $C$ that does not depend on $R$. This allows us to expand the integrand for small $\omega'/R$ in order to obtain the asymptotic behavior at large $R$. In particular, we can approximate
	\begin{equation}
	16^{-R}\left[4+\left(\frac{\omega'}{R}\right)^2-\frac{\omega'}{R}\sqrt{8+\left(\frac{\omega'}{R}\right)^2}\right]^{2R}\approx 16^{-R}\left(4-\frac{\sqrt{8}\omega'}{R}\right)^{2R}= \left(1-\frac{\omega'}{\sqrt{2}R}\right)^{2R}\approx e^{-\sqrt{2}\omega'}.\label{eq:useful_approximation}
	\end{equation}
	Keeping only the leading contributions also in the rest of the integral, we obtain
	\begin{equation}
	J_{\mathrm{RKKY}}^{AA}\approx-\frac{J_1J_2}{8\sqrt{2}\pi R}\int_0^\infty d\omega'\, \omega' e^{-\sqrt{2}\omega'}=-\frac{J_1J_2}{16\sqrt{2}\pi R}.\label{eq:EAAasymptotic}
	\end{equation}
	After reinstating $a$ and $t$, this leads us to Eq.~(\ref{eq:RKKY_asymptotic_AA}). 
	
	In a similar fashion, we can also obtain an asymptotic expression for the $AB$ configuration. Since the second impurity is now located on the $B$ sublattice, this impurity will also affect the flat band. In this case, the vanishing band width of the flat band makes it questionable whether we can expand our expression for the RKKY coupling in orders of $J_2$. We therefore keep the full $T$-matrix for the second impurity while still expanding in orders of $J_1$. To lowest order in $J_1$, we then find two different nonvanishing contributions to the RKKY coupling:
	\begin{align}
	&J_{\mathrm{RKKY}}^{AB,12}=J_{\mathrm{RKKY}}^{AB,21}=-\frac{2J_1J_2}{\pi}\int_0^\infty d\omega\,(-i\omega)\, \frac{G^{(0)}_{BA}(R,i\omega)P^{(0)}_{AB}(-R,i\omega)}{4-J_2^2[G^{(0)}_{BB}(0,i\omega)]^2},\label{eq:sm_JAB1}\\&J_{\mathrm{RKKY}}^{AB,22}=-\frac{4J_1J_2^3}{\pi}\int_0^\infty d\omega\,(-i\omega)\, \frac{G^{(0)}_{BA}(R,i\omega)G^{(0)}_{AB}(-R,i\omega)G^{(0)}_{BB}(0,i\omega)P^{(0)}_{BB}(0,i\omega)}{\left(4-J_2^2[G^{(0)}_{BB}(0,i\omega)]^2\right)^2}.\label{eq:sm_JAB2}
	\end{align}
	Plugging in the Green functions, we get
	\begin{align}
	J_{\mathrm{RKKY}}^{AB,12}&=\frac{8J_1J_2}{\pi}\int_0^\infty d\omega\,\omega\left[\frac{16^{-R}\omega^2(\omega-\gamma)(4+\omega^2-\omega\gamma)^{2R-2}[4+(\omega^2-\omega\gamma)(1-R)-8R]}{\gamma^2[8\omega^2\gamma^2+(4+\omega^2)J_2^2+\omega\gamma J_2^2]}\right],\\
	J_{\mathrm{RKKY}}^{AB,22}&=\frac{4J_1J_2^3}{\pi}\int_0^\infty d\omega\,\omega\left[\frac{16^{-R}\omega(4+\omega^2-\omega\gamma)^{2R}[-8\gamma+\omega(8+\omega^2-2\omega\gamma)]}{\gamma^3[\omega^2\gamma^2(4+\omega^2-\omega\gamma)+2J_2^2]^2}\right].
	\end{align}
	These integrals can be approximated following the same steps as above, and, in particular, using again Eq.~(\ref{eq:useful_approximation}). We obtain
	\begin{equation}
	J_{\mathrm{RKKY}}^{AB}=2J_{\mathrm{RKKY}}^{AB,12}+J_{\mathrm{RKKY}}^{AB,22}\approx \frac{J_1}{\pi J_2R^3}\int_0^\infty d\omega'\,\left(\frac{\omega'^3}{\sqrt{2}}-\omega'^2\right)e^{-\sqrt{2}\omega'} =\frac{J_1}{2\sqrt{2}\pi J_2R^3},
	\end{equation}
	which leads us to Eq.~(\ref{eq:RKKY_asymptotic_AB}). Finally, for the $BB$ configuration, we now keep the full $T$-matrices for both impurities. We now get three different nonvanishing contributions to the RKKY coupling:
	\begin{align}
	&J_{\mathrm{RKKY}}^{BB,12}=J_{\mathrm{RKKY}}^{BB,21}\nonumber\\&=-\frac{8J_1J_2}{\pi}\int_0^\infty d\omega\,(-i\omega)\frac{G^{(0)}_{BB}(R,i\omega)P^{(0)}_{BB}(-R,i\omega)\{16-4G^{(0)}_{BB}(0,i\omega)^2(J_1^2+J_2^2)+[G^{(0)}_{BB}(0,i\omega)^4-G^{(0)}_{BB}(R,i\omega)^4]J_1^2J_2^2\}}{F_1(R,i\omega)F_2(R,i\omega)},\\
	&J_{\mathrm{RKKY}}^{BB,11}=\frac{16J_1^3J_2}{\pi}\int_0^\infty d\omega\,(i\omega)\frac{G^{(0)}_{BB}(0,i\omega)P^{(0)}_{BB}(0,i\omega)G^{(0)}_{BB}(R,i\omega)^2\{4-[G^{(0)}_{BB}(0,i\omega)^2-G^{(0)}_{BB}(R,i\omega)^2]J_2^2\}}{F_1(R,i\omega)F_2(R,i\omega)},\\
	&J_{\mathrm{RKKY}}^{BB,22}=\frac{16J_1J_2^3}{\pi}\int_0^\infty d\omega\,(i\omega)\frac{G^{(0)}_{BB}(0,i\omega)P^{(0)}_{BB}(0,i\omega)G^{(0)}_{BB}(R,i\omega)^2\{4-[G^{(0)}_{BB}(0,i\omega)^2-G^{(0)}_{BB}(R,i\omega)^2]J_1^2\}}{F_1(R,i\omega)F_2(R,i\omega)},
	\end{align}
	with $F_{1,2}(R,i\omega)=16-4G^{(0)}_{BB}(0,i\omega)^2(J_1^2+J_2^2)\pm8G^{(0)}_{BB}(R,i\omega)^2J_1J_2+[G^{(0)}_{BB}(R,i\omega)^2-G^{(0)}_{BB}(0,i\omega)^2]^2J_1^2J_2^2$ and where we have used that $G^{(0)}_{BB}(R,i\omega)=[G^{(0)}_{BB}(R,-i\omega)]^*=G^{(0)}_{BB}(-R,i\omega)$. After plugging in the Green functions, the full expressions become too involved to be displayed here. Nevertheless, the integrals can be approximated in the same way as before, which leads us to Eq.~(\ref{eq:RKKY_asymptotic_BB}):
	\begin{equation}
	J_{\mathrm{RKKY}}^{BB}=2J_{\mathrm{RKKY}}^{BB,12}+J_{\mathrm{RKKY}}^{BB,11}+J_{\mathrm{RKKY}}^{BB,22}\approx\frac{1}{J_1J_2\pi R^5} \int_0^\infty d\omega'\left(-4\sqrt{2}\omega'^5+16\omega'^4\right)e^{-\sqrt{2}\omega'}= -\frac{12\sqrt{2}}{J_1J_2\pi R^5}.
	\end{equation}

	\section{Exact results using Green functions}
	\label{app:GF_numerics}
	
	In this appendix, we present an efficient algorithm that computes the RKKY coupling numerically using the exact lattice Green functions of the full system. For this, we start by noting that the Hamiltonians studied in the main text have a block-tridiagonal structure. The calculation of Green functions for such Hamiltonians, or in general the calculation of inverse matrices of this kind, has been vastly optimized using several methods~\cite{Mueller2020, Meurant1992, Guinea1983, Sancho1985, Lewenkopf2013, Odashima2016, Dy1979}. In this work, we use an algorithm that can efficiently invert block-tridiagonal matrices as described in Ref.~\cite{Mueller2020}.  This algorithm is especially efficient if only a few diagonals, rows, or columns of the matrix are needed since it scales linearly with the system size in these cases. In contrast to Ref.~\cite{Mueller2020} we do not study translationally invariant systems due to the presence of impurities. Therefore, we modify the algorithm of Ref.~\cite{Mueller2020} by adding the needed position dependency as shown in Ref.~\cite{Andergassen2004} for tridiagonal matrices. For convenience we recapitulate the algorithm here and present its modified version.
	
	Since the Hamiltonian is Hermitian, the matrix that needs to be inverted has the following structure:
	\begin{align}
	A=
	\left(
	\begin{array}{cccccc}
	a_1    &   b_1 &  \\
	b_1^\dagger & a_2 & b_2 & \\
	& b_2^\dagger & a_3 & b_3 &\\
	& & b_3^\dagger &  a_4     & \ddots&     \\
	& & & \ddots & \ddots & b_{N-1}\\
	&  &  &  & b_{N-1}^\dagger &  a_N      
	\end{array}
	\right).
	\end{align}
	Here, the $a_n$ and $b_n$ are the blocks that the matrix consists of and $N$ is the number of diagonal blocks. Using a $UDL$-decomposition this matrix can be decomposed into $A=UDL$ with matrices of the form
	\begin{align}
	U =\left(\begin{array}{cccccc}1 & U_{1} & & & & \\ & 1 & U_{2} & & & \\ & & \ddots & \ddots & & \\ & & & \ddots & U_{N-1} \\ & & & & 1\end{array}\right), \hspace{2cm}
	L =\left(\begin{array}{ccccc}1 & & & & \\ L_{1} & 1 & & & \\ & L_{2} & \ddots & & \\ & & \ddots & \ddots & \\ & & & L_{N-1} & 1\end{array}\right),
	\end{align}
	and a block-diagonal matrix $D$ with blocks $D_n$ for $n\in \{1, \ldots ,N\}$. The matrix elements can be calculated using the recursion relations 
	\begin{align}
	D_N &=a_N, \\ 
	U_n &=b_n D_{n+1}^{-1}, \\
	L_n &=D_{n+1}^{-1} b_n^\dagger,\\
	D_n &=a_n-U_n b_n^\dagger \notag \\
	&=a_n-b_n D_{n+1}^{-1} b_n^\dagger \, .
	\end{align}
	The inverse of the matrix $A$ can then be decomposed as well and we find $B=A^{-1}=L^{-1}D^{-1}U^{-1}$ with
	
	\begin{align}
	D^{-1}&=\left(\begin{array}{cccc}
	D_1^{-1} &   \\
	& D_2^{-1}   \\
	& & \ddots \\
	& & &  D_N^{-1} 
	\end{array}\right)\, , \hspace{1cm}
	U^{-1}=\left(\begin{array}{ccccc}
	1 & -U_{1} & U_{1} U_{2} & \cdots & (-1)^{N-1} U_{1} \cdots U_{N-1} \\
	& 1 & -U_{2} & \ddots & \vdots \\
	& & \ddots & \ddots & U_{N-2} U_{N-1} \\
	& & & \ddots & -U_{N-1} \\
	& & & & 1
	\end{array}\right) \, , \notag \\
	L^{-1} &=\left(\begin{array}{cccccc}
	1 & & & & \\
	-L_{1} & \ddots & & \\
	L_{2} L_{1} & \ddots & \ddots & \\
	\vdots & \ddots & -L_{N-2} & 1 & \\
	(-1)^{N-1} L_{N-1} \cdots L_{1} & \cdots & L_{N-1} L_{N-2} & -L_{N-1} & 1
	\end{array}\right) .
	\end{align}
	The diagonal elements of $B$ can be calculated recursively by exploiting the relations
	\begin{align}
	B_{1,1}&=D_1^{-1}, \\
	B_{n+1, n+1}&=D_{n+1}^{-1}+L_n B_{n, n} U_n \notag \\
	&=D_{n+1}^{-1}+D_{n+1}^{-1} b_n^\dagger B_{n, n} b_n D_{n+1}^{-1} \, .
	\end{align}
	The off-diagonal matrix elements can then be computed as well. With $m\geq n$, we find the recursive formulas
	\begin{align}
	B_{n, m+1}&=- B_{n,m} U_m, \label{eq:inversion_bottom}\\
	B_{m+1, n}&=-L_m B_{m, n}, \label{eq:inversion_right}
	\end{align}
	such that we are able to calculate all matrix elements. We want to stress that the $B_{m, n}$ are blocks of the inverse matrix $B$ concerning the different unit cells with elements $B_{m, n}^{\alpha, \beta}$ that can be identified with the Green function $G_{\alpha\beta} (m, n)$.
	
	With these recursion relations we are able to calculate the energy difference, and therefore also the RKKY coupling, by rewriting it as
	\begin{align}
	\Delta E =& \braket{H^{\mathrm{FM}}} - \braket{H^{\mathrm{AFM}}} \\
	=& \sum_{n,m,\alpha,\beta} H_{n\alpha, m\beta}^{\mathrm{FM}} \braket{c_{n, \alpha}^\dagger c_{m,\beta}}_{\mathrm{FM}} - \sum_{n,m,\alpha,\beta} H_{n\alpha, m\beta}^{\mathrm{AFM}} \braket{c_{n, \alpha}^\dagger c_{m,\beta}}_{\mathrm{AFM}} \\
	=& \frac{1}{2 \pi} \int d \omega e^{i \omega 0^{+}}\left[\left(\sum_{n,m,\alpha,\beta} H_{n\alpha, m\beta}^{\mathrm{FM}} G_{\beta \alpha}^{\mathrm{FM}}(m, n; i \omega)\right) -\left(\sum_{n,m,\alpha,\beta} H_{n\alpha, m\beta}^{\mathrm{AFM}} G_{\beta \alpha}^{\mathrm{AFM}}(m, n; i \omega)\right) \right].
	\end{align}
	
	Here, we used that the needed expectation values are given by 
	\begin{align}
	\braket{c_{n,\alpha}^\dagger c_{m, \beta}} &= \frac{1}{i} G_{\beta\alpha}^<(m, n; t=0, t^\prime=0) \\
	&= \frac{1}{2\pi} \int d\omega \, G_{\beta\alpha}^< (m, n; \omega)\\
	&=-\int d\omega \, f(\omega) \left(\frac{1}{\omega - H + i\eta}\right. 
	- \left.\frac{1}{\omega - H - i\eta}\right)_{m\beta,n\alpha} \\
	&= \lim_{T\rightarrow 0} \frac{1}{\beta} \sum_{i\omega_n} G_{\beta\alpha}(m, n; i\omega_n) \, e^{i\omega_n 0^+} \\
	&=\frac{1}{2\pi} \int d\omega \, G_{\beta\alpha}(m, n; i\omega) \, e^{i\omega 0^+} \, .
	\end{align}
	
	For each spin sector this calculation can be done separately. The total energy difference is then given by 
	\begin{align}
	\Delta E_{tot} = \Delta E_\uparrow + \Delta E_\downarrow = 2 \, \Delta E_\uparrow - \frac{J_1+J_2}{2} \, ,
	\end{align}
	assuming that both impurity spins are parallel to spin up in the FM configuration.
	
	\begin{figure}[t]
		\centering
		\includegraphics[width=0.5 \columnwidth]{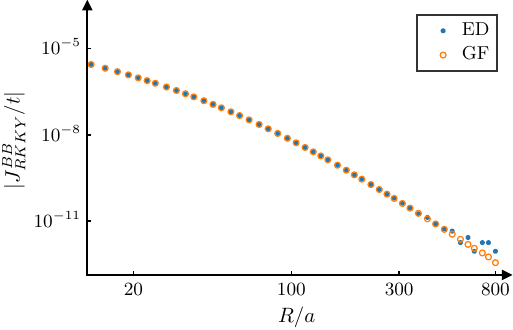}
		\caption{Comparison of the results obtained by exact diagonalization (ED) and the described Green function approach (GF). We show the absolute value of the RKKY coupling $|J_{\mathrm{RKKY}}^{BB}|$ in dependence of the distance $R$ with $J_1=J_2=0.2t$. The results are calculated with a system size of 4000 unit cells but they are converged up to numerical errors. For moderate $R$ we find perfect agreement between the two methods. For very large $R$ we start to see some differences due to the numerical error of the ED. The GF algorithm therefore enables us to calculate the RKKY interaction for larger distances.}
		\label{fig:compare_ED_GF}
	\end{figure}

	Since we only need the first few diagonals of the Green functions to evaluate the formula, the inversion scales only linear with the number of unit cells [$\mathcal{O}(N)$].
	Therefore, the usage of this method allows us to simulate much larger systems (with up to $8 \times 10^5$ unit cells). Additionally, it is possible to calculate the RKKY coupling for larger distances with smaller $J_{\mathrm{RKKY}}$ since the numerical error is reduced. In Fig.~\ref{fig:compare_ED_GF} we compare the results obtained with this algorithm with those calculated with ED.
	We can see that the results for moderate distances perfectly agree. For very large distances we find that the Green functions approach shows even better results than the ED since the numerical error of the ED starts influencing the results.
	
\section{Transition from stub to diamond lattice}
\label{app:transition}
In this appendix, we connect the two models (stub lattice and diamond lattice) by gradually increasing the intercell hopping between the $A$ and $C$ sites from 0 to $t$.

The Hamiltonian is then given by 
\begin{align}
H_{\mathrm{trans}}&=\sum_n\Big(t \, c_{n,A}^\dagger c_{n,B}+ t \, c_{n,A}^\dagger c_{n,C}+t \, c_{n+1,A}^\dagger c_{n,B}+ t^\prime \, c_{n+1,A}^\dagger c_{n,C}\Big)+\mathrm{H.c.}
\label{eq:Hmix}
\end{align}
with $t>0$ and $0\leq t^\prime\leq t$. For $t^\prime=0$ we recover the stub lattice, while $t^\prime=t$ leads to the diamond lattice. In momentum space the corresponding bulk spectrum consists of two dispersive bands $E_\pm(k)=\pm \sqrt{(t^\prime) ^2+3t^2 + 2t(t^\prime+t)\cos(ka)}$ and a third band which always remains flat with $E_0(k)=0$. Therefore, it is of interest to observe how the RKKY interaction changes when $t^\prime$ is changed from $0$ to $t$ so that the gap gradually closes while the flat band remains flat during the transition. In Fig.~\ref{fig:transition} we show the RKKY coupling as a function of the inter-impurity distance for different sublattice configurations. In all cases we find that the RKKY interaction undergoes a crossover from an exponential decay ($t^\prime=0$, stub lattice) to a power-law decay ($t^\prime=t$, diamond lattice). We notice that as $t^\prime$ gets larger and the gap becomes smaller the exponential decay sets in at larger inter-impurity distances until it is pushed to $R\rightarrow\infty$ for $t^\prime\rightarrow t$ and a pure power-law decay is observed for $t^\prime=t$.

	\begin{figure}[t]
		\centering
		\includegraphics[width=\columnwidth]{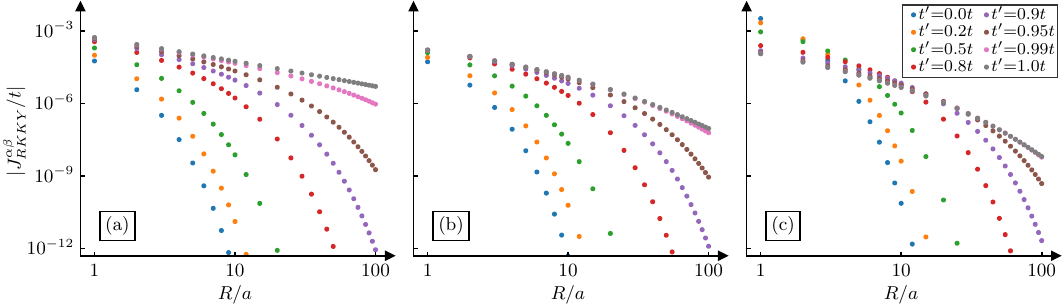}
		\caption{Absolute value of RKKY coupling $|J_{\mathrm{RKKY}}^{\alpha\beta}|$ for $\alpha\beta=$ (a) $AA$, (b) $AB$, and (c) $BB$ in dependence on the inter-impurity distance $R$. The intercell hopping amplitude $t^\prime$ is varied between $0$ (stub lattice) and $t$ (diamond lattice). In (c) the data for $t^\prime=0.99t$ is not visible since it lies below the data for $t^\prime=1.0t$. The RKKY coupling undergoes a crossover from exponential to power-law decay. Here, $J_1=J_2=0.2t$.}
		\label{fig:transition}
	\end{figure}

	\end{widetext}


\begin{thebibliography}{}
		
		\bibitem{Ruderman1954}
		M. A. Rudermann and C. Kittel, Phys. Rev. {\bf 96}, 99 (1954).
		\bibitem{Kasuya1956}
		T. Kasuya, Prog. Theor. Phys. {\bf 16}, 45 (1956).
		\bibitem{Yosida1957}
		K. Yosida, Phys. Rev. {\bf 106}, 893 (1957).
		
		
		\bibitem{Bruno1991}
		P. Bruno and C. Chappert, Phys. Rev. Lett. {\bf 67}, 1602 (1991).
		\bibitem{Bruno1992}
		P. Bruno and C. Chappert, Phys. Rev. B {\bf 46}, 261 (1992).
		
		\bibitem{Craig2004}
		N. J. Craig, J. M. Taylor, E. A. Lester, C. M. Marcus, M. P. Hanson, and A. C. Gossard, Science {\bf 304}, 565 (2004).
		\bibitem{Glazman2004}
		L. I. Glazman and R. C. Ashoori, Science {\bf 304}, 524 (2004).
		\bibitem{Usaj2005}
		G. Usaj, P. Lustemberg, and C. A. Balseiro, Phys. Rev. Lett. {\bf 94}, 036803 (2005).
		\bibitem{Simon2005}
		P. Simon, R. L\'{o}pez, and Y. Oreg, Phys. Rev. Lett. {\bf 94}, 086602 (2005).
		\bibitem{Yang2016}
		G. Yang, C.-H. Hsu, P. Stano, J. Klinovaja, and D. Loss, Phys. Rev. B {\bf 93}, 075301 (2016).
		
		\bibitem{Pientka2013}
		F. Pientka, L. I. Glazman, and F. von Oppen, Phys. Rev. B {\bf 88}, 155420 (2013).
		\bibitem{Braunecker2013}
		B. Braunecker and P. Simon, Phys. Rev. Lett. {\bf 111}, 147202 (2013).
		\bibitem{Klinovaja2013b}
		J. Klinovaja, P. Stano, A. Yazdani, and D. Loss, Phys. Rev. Lett. {\bf 111}, 186805 (2013).
		\bibitem{Vazifeh2013}
		M. M. Vazifeh and M. Franz, Phys. Rev. Lett. {\bf 111}, 206802 (2013).
		\bibitem{Pientka2014}
		F. Pientka, L. I. Glazman, and F. von Oppen, Phys. Rev. B {\bf 89}, 180505(R) (2014).
		\bibitem{Kim2014}
		Y. Kim, M. Cheng, B. Bauer, R. M. Lutchyn, and S. Das Sarma, Phys. Rev. B {\bf 90}, 060401(R) (2014).
		\bibitem{Braunecker2015}
		B. Braunecker and P. Simon, Phys. Rev. B {\bf 92}, 241410(R) (2015).
		\bibitem{Hsu2015}
		C.-H. Hsu, P. Stano, J. Klinovaja, and D. Loss, Phys. Rev. B {\bf 92}, 235435 (2015).
		\bibitem{Schecter2016}
		M. Schecter, K. Flensberg, M. H. Christensen, B. M. Andersen, and J. Paaske, Phys. Rev. B {\bf 93}, 140503(R) (2016).
		\bibitem{Pawlak2016}
		R. Pawlak, M. Kisiel, J. Klinovaja, T. Meier, S. Kawai, T. Glatzel, D. Loss, and E. Meyer, npj Quantum Inf.
		{\bf 2}, 16035 (2016).
		\bibitem{Pawlak2019}
		R. Pawlak, S. Hoffman, J. Klinovaja, D. Loss, and E. Meyer, Progress in Particle and Nuclear Physics {\bf 107}, 1 (2019).
		
		\bibitem{Zyuzin1986}
		A. Y. Zyuzin and B. Z. Spivak, JETP Lett. {\bf 43}, 234 (1986).
		\bibitem{Poilblanc1994}
		D. Poilblanc, D. J. Scalapino, and W. Hanke, Phys. Rev. Lett. {\bf 72}, 884 (1994).
		\bibitem{Balatsky1995}
		A. V. Balatsky, M. I. Salkola, and A. Rosengren, Phys. Rev. B {\bf 51}, 15547 (1995).
		\bibitem{Galitski2002}
		V. M. Galitski and A. I. Larkin, Phys. Rev. B {\bf 66}, 064526 (2002).
		\bibitem{Imamura2004}
		H. Imamura, P. Bruno, and Y. Utsumi, Phys. Rev. B {\bf 69}, 121303(R) (2004).
		\bibitem{Saremi2007} 
		S. Saremi, Phys. Rev. B {\bf 76}, 184430 (2007).
		\bibitem{Hwang2008}
		E. H. Hwang and S. Das Sarma, Phys. Rev. Lett. {\bf 101}, 156802 (2008).
		\bibitem{Braunecker2009}
		B. Braunecker, P. Simon, and D. Loss, Phys. Rev. Lett. {\bf 102}, 116403 (2009).
		\bibitem{Gao2009}
		J. Gao, W. Chen, X. C. Xie, and F.-c. Zhang, Phys. Rev. B {\bf 80}, 241302(R) (2009).
		\bibitem{Liu2009}
		Q. Liu, C.-X. Liu, C. Xu, X.-L. Qi, and S.-C. Zhang, Phys. Rev. Lett. {\bf 102}, 156603 (2009).
		\bibitem{Garate2010}
		I. Garate and M. Franz, Phys. Rev. B {\bf 81}, 172408 (2010).
		\bibitem{Black-Schaffer2010} 
		A. M. Black-Schaffer, Phys. Rev. B {\bf 81}, 205416 (2010).
		\bibitem{Black-Schaffer2010b}
		A. M. Black-Schaffer, Phys. Rev. B {\bf 82}, 073409 (2010).
		\bibitem{Braunecker2010}
		B. Braunecker, G. I. Japaridze, J. Klinovaja, and D. Loss, Phys. Rev. B {\bf 82}, 045127 (2010).
		\bibitem{Chesi2010}
		S. Chesi and D. Loss, Phys. Rev. B {\bf 82}, 165303 (2010).
		\bibitem{Zhu2011}
		J.-J. Zhu, D.-X. Yao, S.-C. Zhang, and K. Chang, Phys. Rev. Lett. {\bf 106}, 097201 (2011).
		\bibitem{Abanin2011}
		D. A. Abanin and D. A. Pesin, Phys. Rev. Lett. {\bf 106}, 136802 (2011).
		\bibitem{Sherafati2011}
		M. Sherafati and S. Satpathy, Phys. Rev. B {\bf 83}, 165425 (2011).
		\bibitem{Kogan2011}
		E. Kogan, Phys. Rev. B {\bf 84}, 115119 (2011).
		\bibitem{Klinovaja2013}
		J. Klinovaja and D. Loss, Phys. Rev. B {\bf 87}, 045422 (2013).
		\bibitem{Power2013}
		S. R. Power and M. S. Ferreira, Crystals {\bf 3}, 49 (2013). 
		\bibitem{Yao2014}
		N. Y. Yao, L. I. Glazman, E. A. Demler, M. D. Lukin, and J. D. Sau, Phys. Rev. Lett. {\bf 113}, 087202 (2014).
		\bibitem{Zyuzin2014}
		A. A. Zyuzin and D. Loss, Phys. Rev. B {\bf 90}, 125443 (2014).
		\bibitem{Efimkin2014}
		D. K. Efimkin and V. Galitski, Phys. Rev. B {\bf 89}, 115431 (2014).
		\bibitem{Schecter2015}
		M. Schecter, M. S. Rudner, and K. Flensberg, Phys. Rev. Lett. {\bf 114}, 247205 (2015).
		\bibitem{Tsvelik2017}
		A. M. Tsvelik and O. M. Yevtushenko, Phys. Rev. Lett. {\bf 119}, 247203 (2017).
		\bibitem{Kurilovich2017}
		V. D. Kurilovich, P. D. Kurilovich, and I. S. Burmistrov, Phys. Rev. B {\bf 95}, 115430 (2017).
		\bibitem{Hsu2017}
		C.-H. Hsu, P. Stano, J. Klinovaja, and D. Loss, Phys. Rev. B {\bf 96}, 081405(R) (2017).
		\bibitem{Hsu2018}
		C.-H. Hsu, P. Stano, J. Klinovaja, and D. Loss, Phys. Rev. B {\bf 97}, 125432 (2018).
		\bibitem{Legg2019}
		H. F. Legg and B. Braunecker, Sci. Rep. {\bf 9}, 17697 (2019).
		\bibitem{Ovando2019}
		O. \'{A}valos-Ovando, D. Mastrogiuseppe, and S. E. Ulloa, Phys. Rev. B {\bf 99}, 035107 (2019).
		\bibitem{Deb2021}
		O. Deb, S. Hoffman, D. Loss, and J. Klinovaja, Phys. Rev. B {\bf 103}, 165403 (2021).
		\bibitem{Laubscher2022}
		K. Laubscher, D. Miserev, V. Kaladzhyan, D. Loss, and J. Klinovaja, Phys. Rev. B {\bf 107}, 115421 (2023).
		
	
		\bibitem{Liu2014}
		Z. Liu, F. Liu, and Y.-S. Wu, Chinese Phys. B {\bf 23}, 077308 (2014).
		\bibitem{Leykam2018}
		D. Leykam, A. Andreanov, and S. Flach, Adv. Phys.: X {\bf 3}, 1473052 (2018).
		
		
		\bibitem{Bistritzer2011}
		R. Bistritzer and A. H. MacDonald, Proc. Natl. Acad. Sci. USA {\bf 108}, 12233 (2011).
		\bibitem{Cao2018}
		Y. Cao, V. Fatemi, A. Demir, S. Fang, S. L. Tomarken, J. Y. Luo, J. D. Sanchez-Yamagishi, K. Watanabe, T. Taniguchi, E. Kaxiras, R. C. Ashoori, and P. Jarillo-Herrero, Nature {\bf 556}, 80 (2018).
		\bibitem{Cao2018b}
		Y. Cao, V. Fatemi, S. Fang, K. Watanabe, T. Taniguchi, E. Kaxiras, P. Jarillo-Herrero, Nature {\bf 556}, 43 (2018).
		
		
		\bibitem{MacDonald2019}
		A. H. MacDonald, Physics {\bf 12}, 12 (2019).
		\bibitem{Andrei2020}
		E. Y. Andrei and A. H. MacDonald, Nat. Mater. {\bf 19}, 1265 (2020).
		\bibitem{Balents2020}
		L. Balents, C. R. Dean, D. K. Efetov, and A. F. Young, Nat. Phys. {\bf 16}, 725 (2020).
		
		
		\bibitem{Lieb1989}
		E. H. Lieb, Phys. Rev. Lett. {\bf 62}, 1201 (1989).
		\bibitem{Sutherland1986}
		B. Sutherland, Phys. Rev. B {\bf 34}, 5208 (1986).
		\bibitem{Mielke1991}
		A. Mielke, J. Phys. A {\bf 24}, L73 (1991).
		\bibitem{Tasaki1992}
		H. Tasaki, Phys. Rev. Lett. {\bf 69}, 1608 (1992).
		\bibitem{Vidal1998}
		J. Vidal, R. Mosseri, and B. Dou\c{c}ot, Phys. Rev. Lett. {\bf 81}, 5888 (1998).
		\bibitem{Mielke1999}
		A. Mielke, Phys. Rev. Lett. {\bf 82}, 4312 (1999).
		\bibitem{Vidal2000}
		J. Vidal, B. Dou\c{c}ot, R. Mosseri, and P. Butaud, Phys. Rev. Lett. {\bf 85}, 3906 (2000).
		
		
		\bibitem{Shen2010}
		R. Shen, L. B. Shao, B. Wang, and D. Y. Xing, Phys. Rev. B {\bf 81}, 041410(R) (2010).
		\bibitem{Apaja2010}
		V. Apaja, M. Hyrk\"{a}s, and M. Manninen, Phys. Rev. A {\bf 82}, 041402(R) (2010).
		\bibitem{Zhang2015}
		T. Zhang and G.-B. Jo, Sci. Rep. {\bf 5}, 16044 (2015).
		\bibitem{Slot2017}
		M. R. Slot, T. S. Gardenier, P. H. Jacobse, G. C. P. van Miert, S. N. Kempkes, S. J. M. Zevenhuizen, C. Morais Smith, D. Vanmaekelbergh, and I. Swart, Nature Phys. {\bf 13}, 672 (2017).
		\bibitem{Xia2018}
		S. Xia, A. Ramachandran, S. Xia, D. Li, X. Liu, L. Tang, Y. Hu, D. Song, J. Xu, D. Leykam, S. Flach, and Z. Chen, Phys. Rev. Lett. {\bf 121}, 263902 (2018).
		\bibitem{Huda2020}
		M. N. Huda, S. Kezilebieke, and P. Liljeroth, Phys. Rev. Research {\bf 2}, 043426 (2020).
			
			
		\bibitem{note3}
		Throughout this work, we use the term `RKKY interaction' to refer to any carrier-mediated indirect exchange interaction between two classical magnetic impurities, not limiting ourselves to the standard RKKY approximation.
		
		
		\bibitem{Bunder2009}
		J. E. Bunder and H.-H. Lin, Phys. Rev. B {\bf 80}, 153414 (2009).
		\bibitem{Cao2019}
		J. Cao, H. A. Fertig, and S. Zhang, Phys. Rev. B {\bf 99}, 205430 (2019).
		\bibitem{Oriekhov2020}
		D. O. Oriekhov and V. P. Gusynin, Phys. Rev. B {\bf 101}, 235162 (2020).
		\bibitem{Bouzerar2021}
		G. Bouzerar, Phys. Rev. B {\bf 104}, 155151 (2021).
		\bibitem{Bouzerar2022}
		G. Bouzerar, Phys. Rev. B {\bf 107}, 184441 (2023).
	
	
		
		\bibitem{Bloembergen1955}
		N. Bloembergen and T. J. Rowland, Phys. Rev. {\bf 97}, 1679 (1955).
		
		\bibitem{Abrikosov1988}
		A. A. Abrikosov, \textit{Fundamentals of the Theory of Metals} (Elsevier, Amsterdam, 1988).
		
		\bibitem{note2}
		Note that, in our model, the unperturbed Hamiltonian does not depend on spin, such that the unperturbed Green functions used here do not carry a spin label.
		
		
		\bibitem{note4}
		In Figs.\ref{fig:RKKY_stub_J}(c) and \ref{fig:RKKY_stub_J}(d) we have $J_{\mathrm{RKKY}}^{AB}\neq J_{\mathrm{RKKY}}^{BA}$ even at $J_1=J_2$ since, for simplicity, the inter-impurity distance $R$ was only defined as the distance between the unit cells in which the impurities are located.
		
		
		
		\bibitem{Economou2006}
		E. N. Economou, \textit{Green's Functions in Quantum Physics}, third edition (Springer, Berlin, 2006).
		
		
		\bibitem{Khalaf2019}
		E. Khalaf, A. J. Kruchkov, G. Tarnopolsky, and A. Vishwanath, Phys. Rev. B {\bf 100}, 085109 (2019).
		\bibitem{Carr2020}
		S. Carr, C. Li, Z. Zhu, E. Kaxiras, S. Sachdev, and A. Kruchkov, Nano Lett. {\bf 20}, 3030 (2020).
		\bibitem{Lei2021}
		C. Lei, L. Linhart, W. Qin, F. Libisch, and A. H. MacDonald, Phys. Rev. B {\bf 104}, 035139 (2021).
		\bibitem{Kim2021}
		H. Kim, Y. Choi, C. Lewandowski, A. Thomson, Y. Zhang, R. Polski, K. Watanabe, T. Taniguchi, J. Alicea, and S. Nadj-Perge, arXiv:2109.12127.
		\bibitem{Shen2022}
		C. Shen, P. J. Ledwith, K. Watanabe, T. Taniguchi, E. Khalaf, A. Vishwanath, and D. K. Efetov, Nat. Mater. {\bf 22}, 316 (2023).
		\bibitem{Li2022}
		Y. Li, S. Zhang, F. Chen, L. Wei, Z. Zhang, H. Xiao, H. Gao, M. Chen, S. Liang, D. Pei, L. Xu, K. Watanabe, T. Taniguchi, L. Yang, F. Miao, J. Liu, B. Cheng, M. Wang, Y. Chen, and Z. Liu, Adv. Mater. {\bf 34}, 2205996 (2022).
		
		\bibitem{Bergholtz2013}
		E. J. Bergholtz and Z. Liu, Int. J. Mod. Phys. B {\bf 27}, 1330017 (2013).
		
		
		
		\bibitem{Mueller2020}
		N. M\"{u}ller, D. M. Kennes, J. Klinovaja, D. Loss, and H. Schoeller, Phys. Rev. B {\bf 101}, 155417 (2020).
		\bibitem{Guinea1983}
		F. Guinea, C. Tejedor, F. Flores, and E. Louis, Phys. Rev. B {\bf 28}, 4397 (1983).
		\bibitem{Sancho1985}
		M. P. Lopez Sancho, J. M. Lopez Sancho, and J. Rubio, J. Phys. F: Met. Phys. {\bf 15}, 851 (1985).
		\bibitem{Lewenkopf2013}
		C. H. Lewenkopf and E. R. Mucciolo, Journal of Computational Electronics {\bf 12}, 203 (2013).
		\bibitem{Odashima2016}
		M. M. Odashima, B. G. Prado, and E. Vernek, Rev. Bras. Ens. Fis. {\bf 39}, e1303 (2017).
		\bibitem{Dy1979}
		K. S. Dy, S.-Y. Wu, and T. Spratlin, Phys. Rev. B {\bf 20}, 4237 (1979).
		
		
		\bibitem{Meurant1992}
		G. Meurant, SIAM J. Matrix Anal. Appl. {\bf 13}, 707 (1992).
		
		\bibitem{Andergassen2004}
		S. Andergassen, T. Enss, V. Meden, W. Metzner, U. Schollw\"ock, and K. Sch\"onhammer, Phys. Rev. B {\bf 70}, 075102 (2004).
		

		
	\end{thebibliography}
\end{document}